\pdfoutput=1

\documentclass[11pt]{article}
\usepackage{graphicx}
\usepackage{amsmath}
\usepackage{amsfonts}
\usepackage{amssymb}
\usepackage[usenames,dvipsnames]{color}

\newcommand{\be}{\begin{equation}}
\newcommand{\ee}{\end{equation}}
\newcommand{\ba}{\begin{eqnarray}}
\newcommand{\ea}{\end{eqnarray}}
\newcommand{\nn}{\nonumber}
\newcommand{\barr}{\begin{array}}
\newcommand{\earr}{\end{array}}

\newcommand\lsim{\mathrel{\rlap{\lower4pt\hbox{\hskip1pt$\sim$}}
        \raise1pt\hbox{$<$}}}
\newcommand\gsim{\mathrel{\rlap{\lower4pt\hbox{\hskip1pt$\sim$}}
        \raise1pt\hbox{$>$}}}

\def\x{{\bf x}}
\def\k{{\bf k}}
\def\fnlcmb{f_{NL}}

\def\Var{\mbox{Var}}
\def\hP{{\hat P}}
\def\bigoh{{\mathcal O}}

\begin{document}

\begin{titlepage}
\setcounter{page}{1} \baselineskip=15.5pt \thispagestyle{empty}

\bigskip\
\begin{center}
{\Large \bf Local stochastic non-Gaussianity and $N$-body simulations}
\end{center}
\vspace{0.5cm}
\begin{center}
{\fontsize{14}{30}\selectfont Kendrick M.~Smith$^1$ and Marilena LoVerde$^2$}
\end{center}

\begin{center}
\vskip 8pt
\textsl{${}^1$ Princeton University Observatory, Peyton Hall, Ivy Lane, Princeton, NJ 08544 USA}
\vskip 4pt
\textsl{${}^2$ Institute for Advanced Study, Einstein Drive, Princeton, NJ 08540, USA}
\end{center}
\vspace{1.2cm}

\hrule \vspace{0.3cm}
{ \noindent \textbf{Abstract} \\[0.2cm]
\noindent
Large-scale clustering of highly biased tracers of large-scale structure has emerged as one of the 
best observational probes of primordial non-Gaussianity of the local type (i.e.~$f_{NL}^{\rm local}$).
This type of non-Gaussianity can be generated in multifield models of inflation such as the curvaton model.
Recently, Tseliakhovich, Hirata, and Slosar showed that the clustering statistics depend qualitatively on
the ratio of inflaton to curvaton power $\xi$ after reheating, a free parameter of the model.
If $\xi$ is significantly different from zero, so that the inflaton makes a non-negligible contribution to
the primordial adiabatic curvature, then the peak-background split ansatz predicts that the halo bias
will be stochastic on large scales.
In this paper, we test this prediction in $N$-body simulations.
We find that large-scale stochasticity is generated, in qualitative agreement with the prediction, but
that the level of stochasticity is overpredicted by $\approx$30\%.
Other predictions, such as $\xi$ independence of the halo bias, are confirmed by the simulations.
Surprisingly, even in the Gaussian case we do not find that halo model predictions for stochasticity 
agree consistently with simulations, suggesting that semi-analytic modeling of stochasticity is generally
more difficult than modeling halo bias.}
 \vspace{0.3cm}
 \hrule

\vspace{0.6cm}
\end{titlepage}

\newpage
\section{Introduction}
\label{sec:intro}

One of the most exciting prospects for cosmology in the near future is the ability to constrain the physics
of inflation \cite{Guth:1980zm,Guth:1982ec,Hawking:1982cz,Starobinsky:1982ee,Bardeen:1983qw,Kamionkowski:1996zd,Seljak:1996gy}, 
thus probing energy scales which are far beyond the reach of accelerator experiments.
The simplest choice of initial conditions, namely stochastic initial fluctuations which are adiabatic, scalar, Gaussian,
and scale-invariant, has been ruled out at the $\approx$3$\sigma$ level.
Current observations are consistent with either a power-law initial power spectrum which is redder than scale invariant
($n_s-1 \approx -0.04$), and marginally consistent with initial conditions which are scale-invariant but contain
contributions from tensor modes ($r\approx 0.2$) \cite{Komatsu:2010fb}.
The next few years will bring a wealth of new data which will sharpen this picture considerably.

Primordial non-Gaussianity has emerged as a particularly powerful probe of inflation
due to the ability to rule out large qualitative classes of models.
For example, there is a theorem \cite{Maldacena:2002vr,Acquaviva:2002ud,Creminelli:2004yq}
which states that in all models of single field inflation whose power spectrum
is nearly scale invariant, the 3-point function $\langle \zeta(\k_1) \zeta(\k_2) \zeta(\k_3) \rangle$ is observationally
indistinguishable from zero in ``squeezed'' triangles (i.e.~$\min(k_i) \ll \max(k_i)$).
However, a detectably large squeezed 3-point function is naturally generated in other models, such as the ekpyrotic scenario
\cite{Creminelli:2007aq,Buchbinder:2007at,Lehners:2007wc,Koyama:2007if}.
Observational constraints on Gaussianity to date have mainly focused on the 3-point function and have parameterized
the deviation from zero by three parameters $f_{NL}^{\rm local}$, $f_{NL}^{\rm equil}$, $f_{NL}^{\rm orthog}$
\cite{Komatsu:2003iq,Babich:2004gb,Creminelli:2005hu,Yadav:2007yy,Meerburg:2009ys,Smith:2009jr,Senatore:2009gt}.
The current WMAP constraints from \cite{Komatsu:2010fb} are:
$f_{NL}^{\rm local} = 32\pm 21$, $f_{NL}^{\rm equil} = 26 \pm 140$, and $f_{NL}^{\rm orthog} = -202\pm 104$
(errors are $1\sigma$).

In this paper, we will focus on local non-Gaussianity and use the notation $\fnlcmb = f_{NL}^{\rm local}$ throughout.
In this case, the initial curvature fluctuation is of the form $\zeta(\x) = \zeta_G(\x) + \frac{3}{5} \fnlcmb(\zeta_G(\x)^2-\langle \zeta_G^2(\x)\rangle)$,
where $\zeta_G$ is a Gaussian field \cite{Salopek:1990jq,Gangui:1993tt,Komatsu:2001rj}.
Recently, this type of non-Gaussianity has been studied extensively in the context of large-scale structure, beginning
with a pioneering paper by Dalal et al.~\cite{Dalal:2007cu}, which showed that scale-dependent halo bias is
generated on large scales.
Subsequently, this prediction has been confirmed and extended, in both analytical and simulation-based studies \cite{Moscardini:1990zh,
Slosar:2008hx,Pillepich:2008ka,Matarrese:2008nc,Afshordi:2008ru,McDonald:2008sc,Desjacques:2008vf,Giannantonio:2009ak,Grossi:2009an,
Verde:2009hy,Reid:2010vc,Xia:2010pe}.
In particular, in \cite{Desjacques:2008vf} an improved expression for the $f_{NL}$ dependence of the bias was introduced (the improved expression
agrees with the original expression from \cite{Dalal:2007cu} in the limit $k\rightarrow 0$).
In \cite{Slosar:2008hx}, the constraint $f_{NL}^{\rm local} = 20 \pm 25$ was obtained from observations of large-scale halo clustering
in SDSS (the measurement is obtained from a variety of tracer objects, but the statistical weight is dominated by the high-$z$ photometric
quasar sample).
One qualitative finding which will be particularly relevant for this paper is that the scale-dependent bias is non-stochastic, in 
the sense that the correlation coefficient between halos in different mass bins is equal to one, after shot noise has been subtracted.

The curvaton model is a two-field model of inflation in which the source of initial curvature fluctuations is
not the inflaton, but a second field $\sigma$ whose contribution to the energy density during inflation is subdominant
\cite{Linde:1996gt,Lyth:2001nq,Lyth:2002my,Dimopoulos:2003az,Sasaki:2006kq,Huang:2008zj}.
Most studies of the curvaton model have only considered the case where the curvaton contribution
to the primordial curvature fluctuation $\zeta$ is much larger than the inflaton contribution.  In this case, the curvaton field can give
rise to non-Gaussianity of the local type at a detectable level ($\fnlcmb$ is essentially a free parameter of the model).
Recently, Tseliakhovich, Hirata, and Slosar considered the more general case in which the ratio $\xi$ of inflaton and
curvaton contributions to the primordial curvature fluctuation $\zeta$ can be significantly different from zero \cite{Tseliakhovich:2010kf}
(see also \cite{Langlois:2004nn}).
Applying the same theoretical arguments which predict scale-dependent, non-stochastic halo bias for $\xi=0$, the authors argue that
scale-dependent {\em stochastic} halo bias should be present in the more general case where $\fnlcmb$ and $\xi$ are both nonzero.
A two-component hybrid model with similar observational signatures was studied in \cite{Byrnes:2008zy}; in this case $\xi$ is typically of order one
if the non-Gaussianity is large enough to be detectable.

The purpose of this paper is to analyze $N$-body simulations whose initial conditions contain curvaton and inflaton contributions, 
and study the dependence of halo clustering and halo stochasticity on the parameters
$\{ f_{NL}, \xi \}$ of the model.
Although our main interest is the case $\xi\ne 0$, we also present results for the curvaton model with $\xi=0$ as a baseline for comparison.

Throughout this paper we use the WMAP5+BAO+SN fiducial cosmology \cite{Dunkley:2008ie}, with
baryon density $\Omega_bh^2 = 0.0226$, CDM density $\Omega_ch^2 = 0.114$, Hubble parameter $h=0.70$,
spectral index $n_s=0.961$, optical depth $\tau = 0.080$, and power-law initial curvature power spectrum 
$k^3 P_\zeta(k) / 2\pi^2 = \Delta_\zeta^2 (k/k_{\rm piv})^{n_s-1}$ where $\Delta_\zeta^2 = 2.42 \times 10^{-9}$
and $k_{\rm piv} = 0.002$ Mpc$^{-1}$.
All power spectra and transfer functions have been computed using CAMB \cite{Lewis:1999bs}.

\section{Curvaton model with $\xi\ne 0$}

\subsection{Initial conditions}

In this subsection we review the curvaton model, in the same generality as \cite{Tseliakhovich:2010kf}.

The curvaton is assumed to decay before dark matter freezout, so that no dark matter isocurvature mode is generated, and the adiabatic
curvature fluctuation $\zeta$ is a sum of inflaton and curvaton contributions:
\be
\zeta = \zeta_i + \zeta_c
\ee
We assume that the fields $\zeta_i$ and $\zeta_c$ are uncorrelated and that their power spectra $P_{\zeta_i}, P_{\zeta_c}$, are 
proportional, so that we can define a parameter $\xi = (P_{\zeta_i} / P_{\zeta_c})^{1/2}$ which is independent of scale.  
The power spectra of $\zeta_i, \zeta_c$ are thus related to the power spectrum $P_\zeta$ of the total curvature fluctuation by:
\ba
P_{\zeta_i}(k) &=& \frac{\xi^2}{1+\xi^2} P_\zeta(k) \label{eq:pk_inflaton} \\
P_{\zeta_c}(k) &=& \frac{1}{1+\xi^2} P_\zeta(k) \label{eq:pk_curvaton}
\ea
The power spectrum $P_\zeta$ is taken to be of power-law form $(k^3/2\pi^2) P_\zeta(k) = \Delta_\zeta^2 (k/k_{\rm piv})^{n_s-1}$
with parameters $\Delta_\zeta$, $k_{\rm piv}$ given in \S\ref{sec:intro}.

We assume that $\zeta_i$ is a Gaussian field, but $\zeta_c$ is a non-Gaussian field of ``local type'', i.e.
\be
\zeta_c(\x) = \zeta_{c,G}(\x) + \frac{3}{5} \fnlcmb (1 + \xi^2)^2 \left(\zeta_{c,G}^2(\x)-\langle\zeta_{c,G}^2(\x)\rangle\right)  \label{eq:fnlcmb_def}
\ee
where $\zeta_{c,G}$ is a Gaussian field.
Non-Gaussianity of local type is generated if the curvaton potential $V(\sigma)$ is assumed quadratic in $\sigma$.
Throughout this paper, we will take $\{ \fnlcmb, \xi \}$ to be the parameters of the curvaton model.

To get some intuition for this parameterization, it is useful to note that the power spectrum, bispectrum, and 
{\em connected} higher-point functions of the initial curvature fluctuation $\zeta$ depend on $\fnlcmb$ and $\xi$ 
as follows:
\ba
\label{eq:2pt}
\langle \zeta(\k_1) \zeta(\k_2) \rangle &=& P_\zeta(k_1) (2\pi)^3 \delta^3(\k_1+\k_2) + \bigoh(\fnlcmb^2) \\
\label{eq:3pt}
\langle \zeta(\k_1) \zeta(\k_2) \zeta(\k_3) \rangle &=&\frac{3}{5} \fnlcmb B(k_1,k_2,k_3)\delta^3(\k_1+\k_2+\k_3) + \bigoh(\fnlcmb^3) \\
\langle \zeta(\k_1) \zeta(\k_2) \cdots \zeta(\k_N) \rangle_{\rm conn} &=& \Big(\frac{3}{5}\fnlcmb\Big)^{N-2} (1 + \xi^2)^{N-3} F(\k_1,\ldots,\k_N) \delta^3\Big(\sum_i \k_i \Big)+ \bigoh(\fnlcmb^N)\,.
\label{eq:npoints}
\ea
To lowest order in $\fnlcmb$, the 3-point function is proportional to $\fnlcmb$, with no $\xi$ dependence\footnote{We have defined $\fnlcmb$ in Eq.~(\ref{eq:fnlcmb_def})
with the extra factor of $(1+\xi^2)$ so that the 3-point function will have this property.  The parameter $\tilde f_{NL}$ from \cite{Tseliakhovich:2010kf}
is related to our parameterization by $\tilde f_{NL} = \fnlcmb (1 + \xi^2)^2$.}.
This makes it easy to interpret the CMB bispectrum
constraint from \cite{Komatsu:2010fb} as a constraint $\fnlcmb = 32 \pm 21$ ($1\sigma$ error), with no constraint on $\xi$.
The CMB trispectrum constraint from \cite{Smidt:2010sv} can similarly be interpreted as a constraint $\tau_{NL} = (1.35\pm 0.98) \times 10^4$
on the combination of parameters $\tau_{NL} = (\frac{6}{5} \fnlcmb)^2 (1+\xi^2)$,
with the caveat that the bispectrum and trispectrum estimators are not statistically independent and
there are subtleties in combining the assoicated parameter constraints \cite{Creminelli:2006gc}.

The effect of nonzero $\xi$ is to boost the amplitude of the $N$-point correlation functions (where $N\ge 4$) relative to the
amplitude of the 3-point function.
One interesting consequence is that the 4-point function can be made large while keeping the 3-point function within observational
limits on $\fnlcmb$.
(In fact, halo stochasticity, or ``boosting'' the amplitude of the halo-halo power spectrum $P_{hh}$ relative to the amplitude of 
the matter-halo power spectrum $P_{mh}$, can be viewed
as a formal consequence of boosting the primordial 4-point function relative to the 3-point function.)

\subsection{Halo clustering and the peak-background split}
\label{ssec:halo_model}

The peak-background split formalism is a heuristic argument for predicting correlation functions in which
one or more scales is large compared to the scales which are relevant for spherical collapse \cite{Cole:1989vx,Bardeen:1985tr,Mo:1995cs,Sheth:1999mn,Sheth:1999su}.
This formalism can be applied to study non-Gaussian halo clustering in the curvaton model \cite{Slosar:2008hx,Tseliakhovich:2010kf}.
We will review this calculation here, and extend it by including 1-halo terms which will
be relevant for the stochasticity results to be presented later.

Let us write the inflaton contribution $\zeta_i$ to the initial curvature fluctuation as a sum of long-wavelength 
and short-wavelength pieces: $\zeta_i = (\zeta_{i,l} + \zeta_{i,s})$.
We analogously write the Gaussian field $\zeta_{c,G}$ as a sum $(\zeta_{c,l} + \zeta_{c,s})$.
The total initial curvature fluctuation is then given by
\ba
\zeta(\x) &=& \zeta_i(\x) + \zeta_{c,G}(\x) + \frac{3}{5} \fnlcmb (1+\xi^2)^2 \left(\zeta_{c,G}(\x)^2-\langle\zeta_{c,G}(\x)^2\rangle\right) \\
  &=& \zeta_{i,l}(\x) + \zeta_{c,l}(\x) \nn \\
   && \hspace{0.5cm} + \zeta_{i,s}(\x) + \left(1 + \frac{6}{5} \fnlcmb (1+\xi^2)^2 \zeta_{c,l}(\x) \right) \zeta_{c,s}(\x) \nn \\
   && \hspace{0.5cm} + \frac{3}{5} \fnlcmb (1+\xi^2)^2 (\zeta_{c,l}^2 + \zeta_{c,s}^2-\langle\zeta_{c,l}^2\rangle + \langle\zeta_{c,s}^2\rangle)  \label{eq:zeta_tot}
\ea
and we have assumed the long and short wavelength parts of $\zeta_{c}$ are uncorrelated.  
Let us make the approximation that the terms in the third line of~(\ref{eq:zeta_tot}) are negligible.
(The $\zeta_{c,l}^2$ term will not be important for our purposes since we will only use the small-scale component of Eq.~(\ref{eq:zeta_tot});
the main effect of the $\zeta_{c,s}^2$ term is to change the constant part of the halo bias, which is a free parameter anyway.)
The first line is the long-wavelength part of $\zeta$, which is unchanged from the Gaussian case (i.e.~it does not depend on $\fnlcmb$).
The second line is the short-wavelength part; we find that the effect of the non-Gaussianity is to modulate the small-scale 
curvaton mode $\zeta_{c,s}$ by a factor $(1 + \frac{6}{5} \fnlcmb (1+\xi^2)^2 \zeta_{c,l})$ which depends on the long-wavelength
curvaton mode $\zeta_{c,l}$.

In the peak-background split picture, we interpret Eq.~(\ref{eq:zeta_tot}) as saying that the small-scale matter power
spectrum is no longer spatially constant in a non-Gaussian cosmology, but rather a local quantity which varies with position.
If we consider a large box at position $\x$, then the average small-scale power spectrum in the box is given by
\ba
P_\zeta(\x) &=& P_{\zeta_i} + \left(1 + \frac{6}{5} \fnlcmb (1+\xi^2)^2 \zeta_{c,l}(\x) \right)^2 P_{\zeta_c}+ \bigoh(\fnlcmb^2) \nn \\
   & \approx & \left(1 + \frac{6}{5} \fnlcmb (1+\xi^2) \zeta_{c,l}(\x) \right)^2 P_\zeta \label{eq:Pbox}\,.
\ea
Following notation from \cite{Slosar:2008hx}, we will parameterize the amplitude of the small-scale power spectrum
by $\sigma_8$, the RMS of the {\em linear} density field at $z=0$ with $8h^{-1}$ Mpc tophat smoothing, and rephrase
Eq.~(\ref{eq:Pbox}) by writing $\sigma_8$ as a function of position $\x$:
\be
\sigma_8(\x) = \left(1 + \frac{6}{5} \fnlcmb (1+\xi^2) \zeta_{c,l}(\x) \right) \sigma_8  \label{eq:sigma8box}\,.
\ee
Now let us ask how the number density $n_h$ of halos varies on large scales in the peak-background split picture.
The density of halos $n_h(\x)$ in a large box at position $\x$ will differ from the mean density $\bar n_h$ for two reasons:
first, because the local matter density $\rho_m (1 + \delta_l(\x))$ in the box differs from the mean $\rho_m$,
and second because the local value of $\sigma_8$ differs from the mean via Eq.~(\ref{eq:sigma8box}).  Combining
these effects we can write:
\be
n_h(\x) = \bar n_h (1 + \delta_l(\x)) \left( 1 + \delta_l(\x) \frac{\partial \log \bar n_h}{\partial \delta_l} 
  + \frac{6}{5} \fnlcmb (1 + \xi^2) \zeta_{c,l}(\x) \frac{\partial \log \bar n_h}{\partial \log \sigma_8} \right) \label{eq:pbs1}
\ee
The $(1+\delta_l)$ prefactor comes from converting Lagrangian to Eulerian space.
The second term is proportional to the (scale-independent) derivative $(\partial \log \bar n_h / \partial \delta_l)$ of the
mass function with respect to the background density, i.e.~the Lagrangian halo bias.
These two terms are present in a Gaussian cosmology and represent the usual halo bias which is constant on large scales.
The third term represents the effects of primordial non-Gaussianity, which gives an extra scale-dependent contribution.

Taking the Fourier transform of Eq.~(\ref{eq:pbs1}) and dropping second-order terms, we get
\ba
\delta_h(\k) &=& \left(1 + \frac{\partial \log \bar n_h}{\partial \delta_l} \right) \delta(\k)
                  + \frac{6}{5} \fnlcmb (1 + \xi^2) \frac{\partial \log \bar n_h}{\partial \log \sigma_8} \zeta_c(\k) \nn \\
  &=& b_G \delta(\k) + (1+\xi^2) b_{NG}(k) \delta_c(\k)  \label{eq:pbs_h}
\ea
where $\delta_h(\k) = n_h(\k)/\bar n_h$ is the fractional halo overdensity, and in the last line we have defined
\ba
\label{eq:bGdef}
b_G &=& 1 + \frac{\partial \log \bar n_h}{\partial \delta_l}  \\
b_{NG}(k) &=& \frac{2 \fnlcmb}{\alpha(k,z)} \frac{\partial \log \bar n_h}{\partial \log \sigma_8}
\ea
where
\be
\alpha(k,z) = \frac{2 k^2 T(k) D(z)}{3 \Omega_m H_0^2}
\ee
The quantity $\alpha(k,z)$ relates the matter overdensity $\delta_m(\k,z)$ to the initial curvature $\zeta(\k)$ in
linear perturbation theory: $\delta_m(\k,z) =\frac{3}{5} \alpha(k,z) \zeta(\k)$.
Note that we have defined inflaton and curvaton contributions to the matter density by
$\delta_i(\k,z) = \frac{3}{5} \alpha(k,z) \zeta_i(\k)$ and
$\delta_c(\k,z) = \frac{3}{5} \alpha(k,z) \zeta_c(\k)$,
even though strictly speaking, Poisson's equation applies only to the sum of the two fields.

The peak-background split expression~(\ref{eq:pbs_h}) for $\delta_h(\k)$ applies on scales which are large compared
to scales relevant for spherical collapse.
It is also incomplete, in the sense that it treats the halo overdensity as a continuous field, and ignores stochastic variations
due to random halo locations.
If we compute matter-halo and halo-halo power spectra using this expression, then we get:
\ba
P^{2H}_{mh}(k) &=& [ b_G + b_{NG}(k) ] P_{\rm lin}(k) \\
P^{2H}_{hh'}(k) &=& \left[ ( b_G + b_{NG}(k) ) ( b'_G + b'_{NG}(k) ) + \xi^2 b_{NG}(k) b'_{NG}(k) \right] P_{\rm lin}(k)
\ea
where the primes refer to different halo masses and $P_{\rm lin}(k)=\frac{9}{25}\alpha^2(k)P_\zeta(k)$ is the linear theory
matter power spectrum.
We have included the superscript ``2H'' because these expressions omit 1-halo terms.
Using standard machinery from the halo model \cite{Seljak:2000gq,Scoccimarro:2000gm,Ma:2000ik,Peacock:2000qk},
it is straightforward to calculate 1-halo contributions to these power spectra.
In non-overlapping mass bins, let $n_i$ be the number density of halos in the $i$-th bin, and
let $f_i$ denote the total fraction (by mass) of dark matter in halos in mass bin $i$.  Then we get:\footnote{These expressions neglect 
convolution by the halo density profiles, but this can be neglected for purposes of this paper, where we only study clustering 
on large scales ($k \lsim 0.04$ $h$ Mpc$^{-1}$).}
\ba
P_{mm}(k) &=& P_{\rm lin}(k) + P_{mm}^{1H}  \label{eq:pmm_1h_2h} \\
P_{mi}(k) &=& (b^{(i)}_G + b^{(i)}_{NG}(k)) P_{\rm lin}(k) + \frac{f_i}{n_i} \\
P_{ij}(k) &=& \Big[ (b^{(i)}_G + b^{(i)}_{NG}(k))(b^{(j)}_G + b^{(j)}_{NG}(k)) + \xi^2 b^{(i)}_{NG}(k) b^{(j)}_{NG}(k) \Big] P_{\rm lin}(k) + \frac{\delta_{ij}}{n_i} \label{eq:phh_1h_2h}
\ea
where we have defined
\be
P_{mm}^{1H} = \rho_m^{-2} \int dM\, M^2 n(M)\,.
\ee
Although the 1-halo terms are generally smaller than the 2-halo terms on the angular scales we will study in this paper ($k \le 0.04$ $h$~Mpc$^{-1}$),
they are the leading souce of stochasticity (aside from shot noise) predicted by the halo model in the Gaussian case.
In the non-Gaussian case, the term $(\xi^2 b^{(i)}_{NG} b^{(j)}_{NG})$ represents extra stochasticity on large scales for nonzero $\fnlcmb$ and $\xi > 0$.
This source of stochasticity can be understood intuitively: the non-Gaussian part of the bias traces the curvaton field $\delta_c$ on large scales, and
if $\xi > 0$, this field is not 100\% correlated to the matter overdensity $\delta$.

There is a useful simplification to the above expression if we assume a ``universal'' halo mass function of the form $dn/dM = (\rho_m/M) f(\nu) (d\nu/dM)$.
Here, $\nu(M,z) = (\delta_c^2/\sigma^2(M,z))$, where $\delta_c = 3 (12\pi)^{2/3}/20 \approx 1.69$ is the threshhold for spherical collapse,
and $\sigma^2(M,z)$ is the variance of the linear matter overdensity after tophat smoothing on the scale corresponding to halo mass $M$.
For a universal mass function, the relation
\be
\frac{\partial \log \bar n}{\partial \log \sigma_8} = \delta_c \frac{\partial \log \bar n}{\partial \delta_l}
\ee
holds \cite{Slosar:2008hx}, so that $b_{NG}$ and $b_G$ are related by
\be
b_{NG}(k) = \frac{2 \delta_c}{\alpha(k,z)} \fnlcmb (b_G - 1)\,.  \label{eq:bng_universal}
\ee
Mass functions obtained from simulations are roughly universal \cite{Press:1973iz,Sheth:1999mn,Jenkins:2000bv,Reed:2006rw},
and we will generally assume that Eq.~(\ref{eq:bng_universal}) holds throughout the paper.

\section{$N$-body simulations}

To study halo clustering in the curvaton model, we
performed collisionless $N$-body simulations using the GADGET-2 TreePM code \cite{Springel:2005mi}.
Simulations were done using periodic
box size $R_{\rm box} = 1600$ $h^{-1}$~Mpc, particle count $N_p = 1024^3$, and force softening
length $R_s = 0.05 (R_{\rm box}/N_p^{1/3})$.
With these parameters and the fiducial cosmology from \S\ref{sec:intro}, the particle mass 
is $m_p = 2.92 \times 10^{11}$ $h^{-1}$~$M_\odot$.
Results in this paper were obtained from two simulations with $\fnlcmb=0$, and two simulations for each choice of
$\fnlcmb \in \{\pm 250, \pm 500\}$ and $\xi \in \{0,1\}$ (for a total of 18 simulations). 

For given curvaton model parameters $\fnlcmb,\xi$, we simulate initial conditions as follows.
First, we simulate Gaussian fields $\zeta_i$ and $\zeta_{c,G}$ in Fourier space with power spectra given by 
Eqs.~(\ref{eq:pk_inflaton}),~(\ref{eq:pk_curvaton}).
We then compute the non-Gaussian curvaton field in real space by $\zeta_c = \zeta_{c,G} + \frac{3}{5} \fnlcmb (1+\xi^2)^2 (\zeta_{c,G}^2-\langle\zeta_{c,G}^2\rangle)$.
(When generating the initial conditions, all Fourier transforms are computed on a grid with $N_p^3$ elements.)
We apply the transfer function $T(k)$ to the total curvature fluctuation
$\zeta = \zeta_i + \zeta_c$ to obtain the
Newtonian potential $\Phi(k)$ at the initial redshift $z_{\rm ini}=100$ of the simulations.
Finally, we obtain initial particle positions using the Zeldovich approximation \cite{Zeldovich:1969sb}.
(At $z_{\rm ini}=100$, transient effects due to use of this approximation should be negligible \cite{Crocce:2006ve}.)

We group particles into halos using an MPI parallelized implementation of the 
friends-of-friends (FOF) algorithm \cite{Frenk:1988zz}
with link length $L_{\rm FOF} = 0.2 R_{\rm box} N_p^{-1/3}$.
For a halo containing $N_{\rm FOF}$ particles, we assign a halo position given by the mean of the individual
particle positions, and a halo mass given by:
\be
m_h = m_p \left( N_{\rm FOF} - N_{\rm FOF}^{0.4} \right)\,.
\ee
The second term is recommended in \cite{Warren:2005ey} to minimize particle resolution artifacts when
estimating the mass function using a FOF halo finder.

In this paper we will analyze matter and halo power spectra, using redshifts and halo mass bins defined in Tab.~\ref{tab:mass_bins}.
We estimate power spectra by assigning particle positions (or halo positions) to a real-space grid with $1536^3$ points using the
Cloud-in-Cell algorithm, taking the Fourier transform, and averaging the power over all Fourier modes 
in a $k$-bin.  This is described in more detail in Appendix~\ref{app:estimators}.

\begin{table}[!h]
\begin{center}
\begin{tabular}{|c|c|c|c|c|}
\hline & Mass range ($h^{-1} M_\odot$) & $n$ ($h^3$ Mpc$^{-3}$) & $f$ & $b_G$ \\ \hline\hline
$z=2$ & $M > 1.15 \times 10^{13}$ & $3.782 \times 10^{-5}$ & 0.010 & $5.222 \pm 0.041$ \\
\hline $z=1$ & $1.15 \times 10^{13} < M < 2.32\times 10^{13}$ & $1.163 \times 10^{-4}$ & 0.024 & $2.502 \pm 0.014$ \\
      & $M > 2.32 \times 10^{13}$ & $6.346 \times 10^{-5}$ & 0.038 & $3.470 \pm 0.018$ \\
\hline $z=0.5$ & $1.15 \times 10^{13} < M < 2.32\times 10^{13}$ & $1.678 \times 10^{-4}$ & 0.035 & $1.720 \pm 0.010$ \\
        & $2.32 \times 10^{13} < M < 4.66\times 10^{13}$ & $7.585 \times 10^{-5}$ & 0.032 & $2.101 \pm 0.014$ \\
        & $M > 4.66 \times 10^{13}$ & $4.498 \times 10^{-5}$ & 0.057 & $2.996 \pm 0.016$ \\
\hline $z=0$ & $1.15 \times 10^{13} < M < 2.32\times 10^{13}$ & $2.020 \times 10^{-4}$ & 0.042 & $1.202 \pm 0.008$ \\
      & $2.32 \times 10^{13} < M < 4.66\times 10^{13}$ & $1.011 \times 10^{-4}$ & 0.043 & $1.437 \pm 0.010$ \\
      & $4.66 \times 10^{13} < M < 1.02\times 10^{14}$ & $5.189 \times 10^{-5}$ & 0.045 & $1.783 \pm 0.014$ \\
      & $M > 1.02\times 10^{14}$ & $2.778 \times 10^{-5}$ & 0.079 & $2.628 \pm 0.015$ \\
\hline
\end{tabular}
\end{center}
\caption{Mass bins used throughout this paper, with halo number density $n$, total fraction (by mass) $f$ of dark matter
in halos in each bin, and Gaussian bias $b_G$ estimated from simulation.
The fitting procedure used to estimate $b_G$ and assign statistical errors is described in \S\ref{ssec:halo_bias} and uses only wavenumbers $k \le 0.04$ $h$~Mpc$^{-1}$.}
\label{tab:mass_bins}
\end{table}

\section{Halo bias}
\label{ssec:halo_bias}

For a halo mass bin $i$, we define the bias parameter
\be
b_{mi}(k) = \frac{P_{mi}(k)}{P_{mm}(k)}\,.
\ee
In this section, we will compare values of $b_{mi}(k)$ estimated from simulation with the predicted form:
\be
b_{mi}(k) = b_0 + \frac{2 \delta_c}{\alpha(k,z)} \fnlcmb (b_0-1)\,.  \label{eq:bias_prediction}
\ee
In writing down this predicted form, we have omitted 1-halo terms derived previously in Eqs.~(\ref{eq:pmm_1h_2h})--(\ref{eq:phh_1h_2h}).
We find that including 1-halo terms does not qualitatively affect any conclusions from this section, but the simplified 
form in Eq.~(\ref{eq:bias_prediction}) is convenient for comparison with the rest of the literature.

\begin{figure}[!ht]
\centerline{
\includegraphics[width=5.7cm]{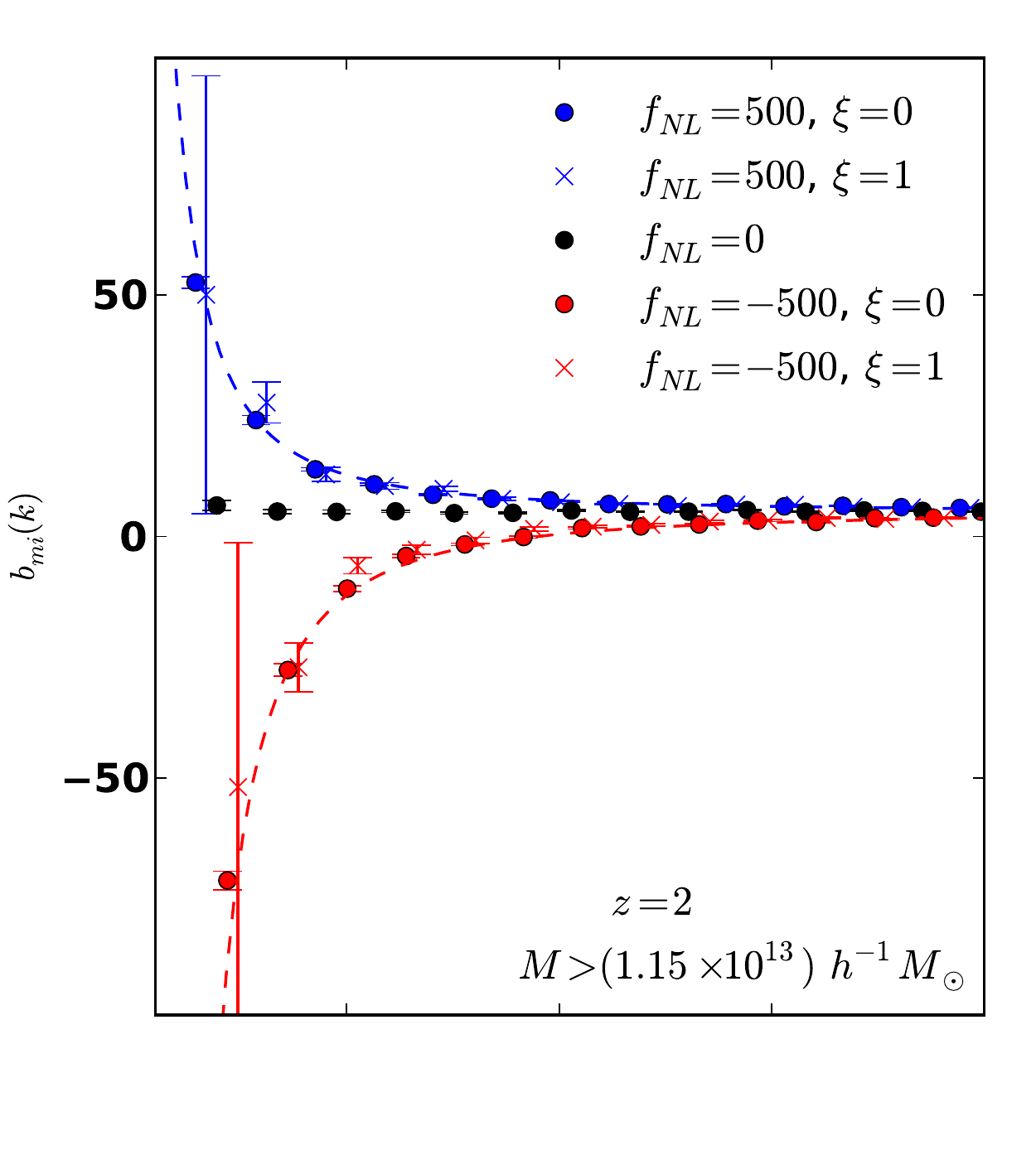}
\includegraphics[width=5.7cm]{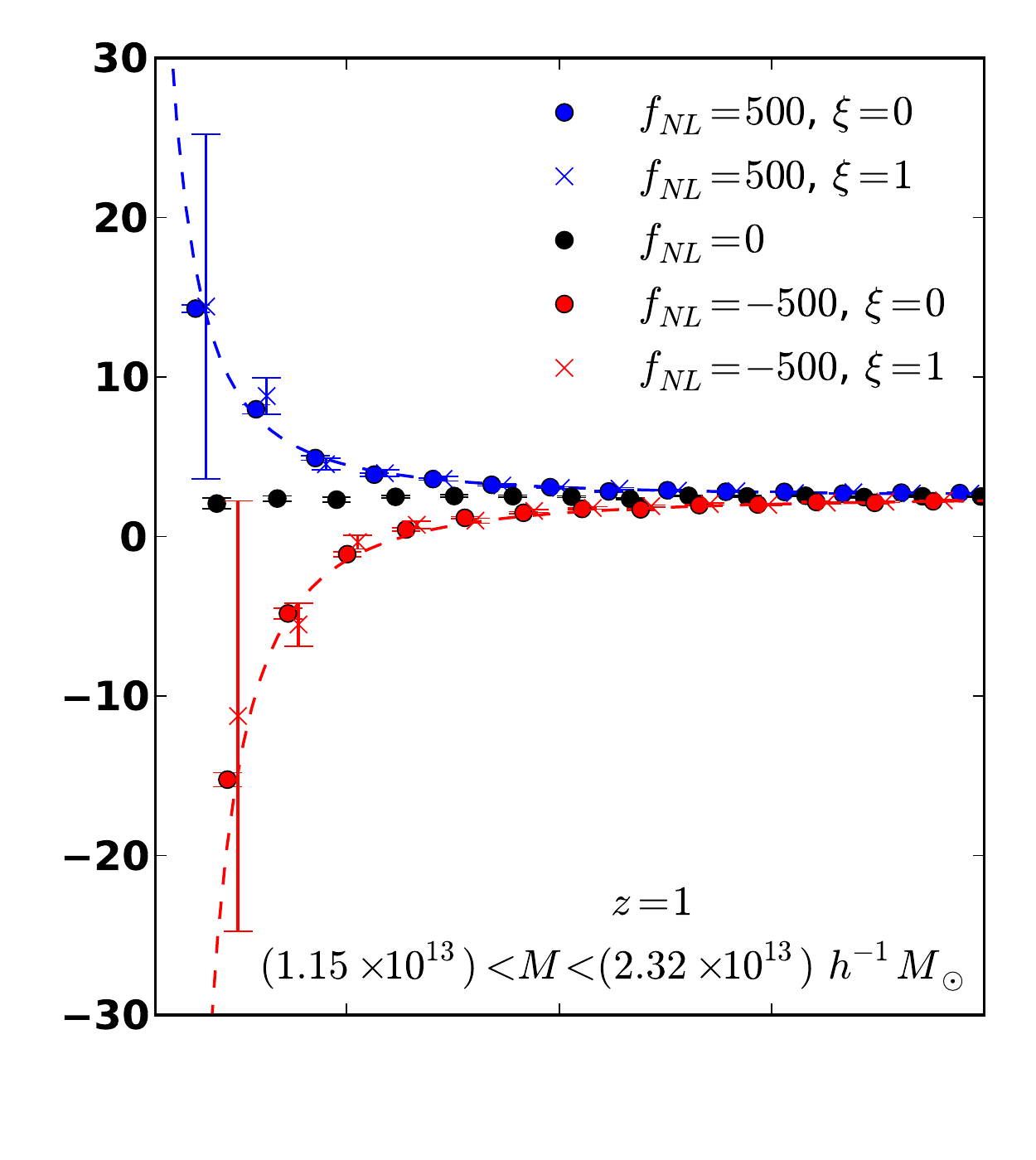}
\includegraphics[width=5.7cm]{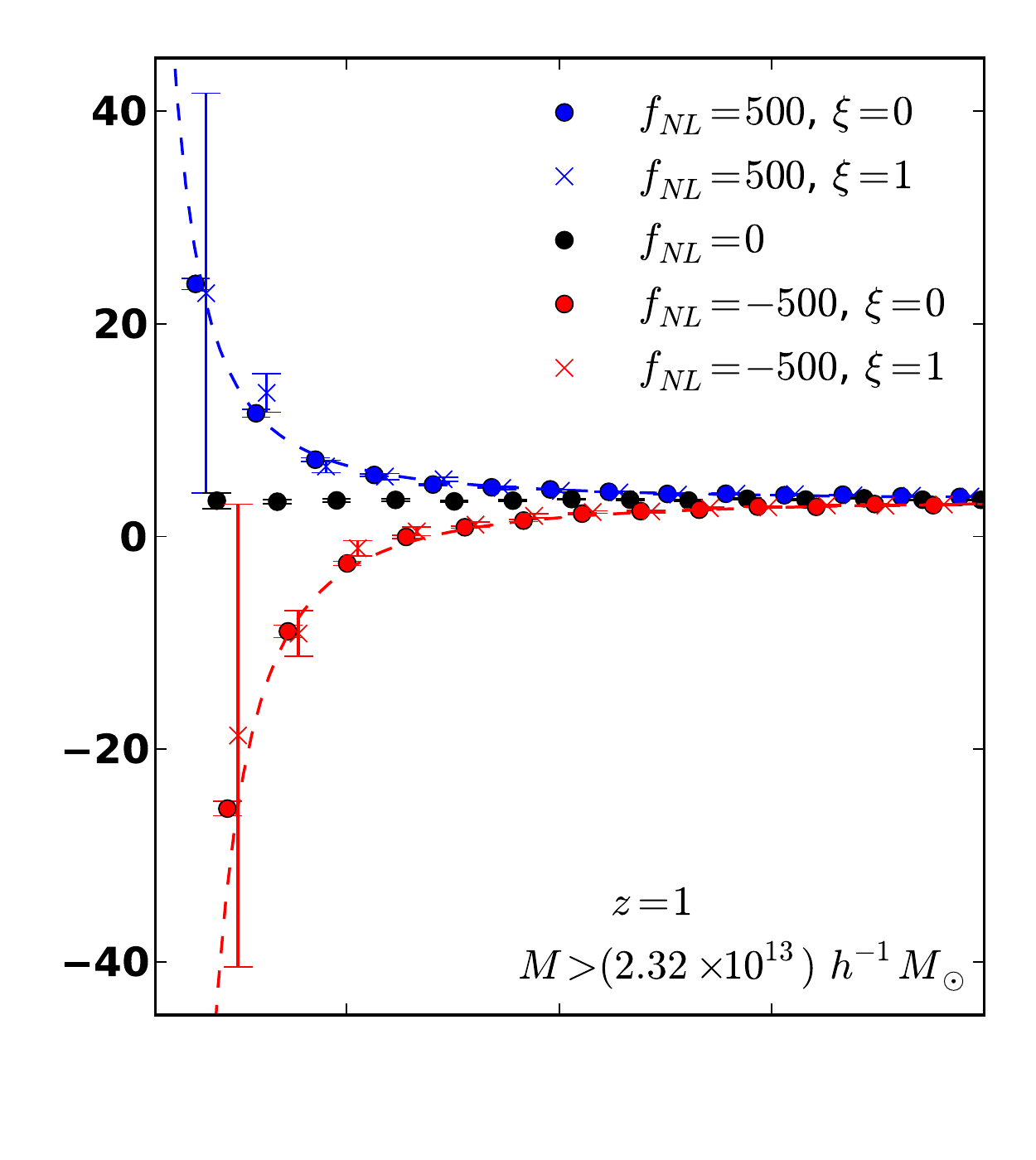}
}
\centerline{
\includegraphics[width=5.7cm]{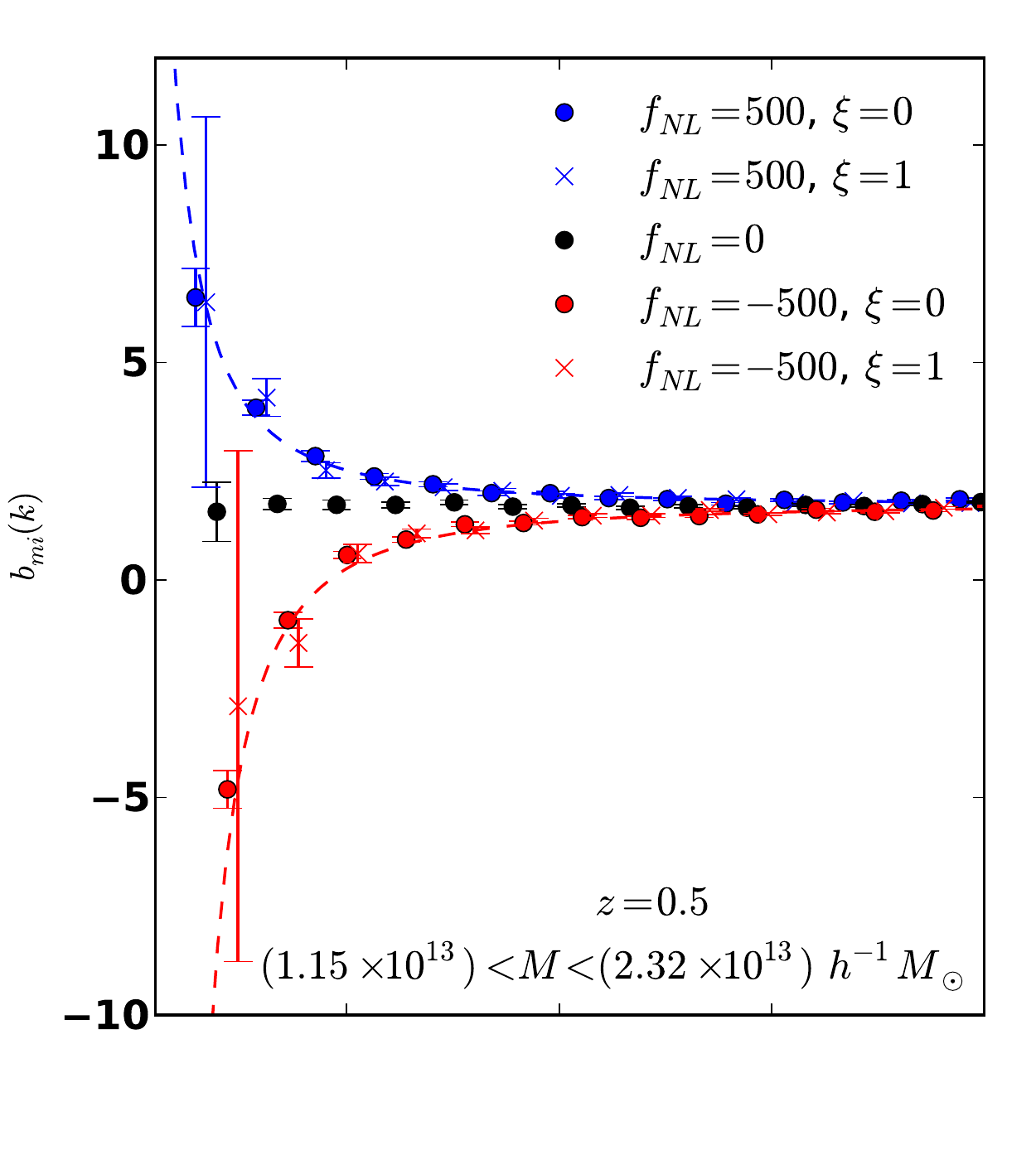}
\includegraphics[width=5.7cm]{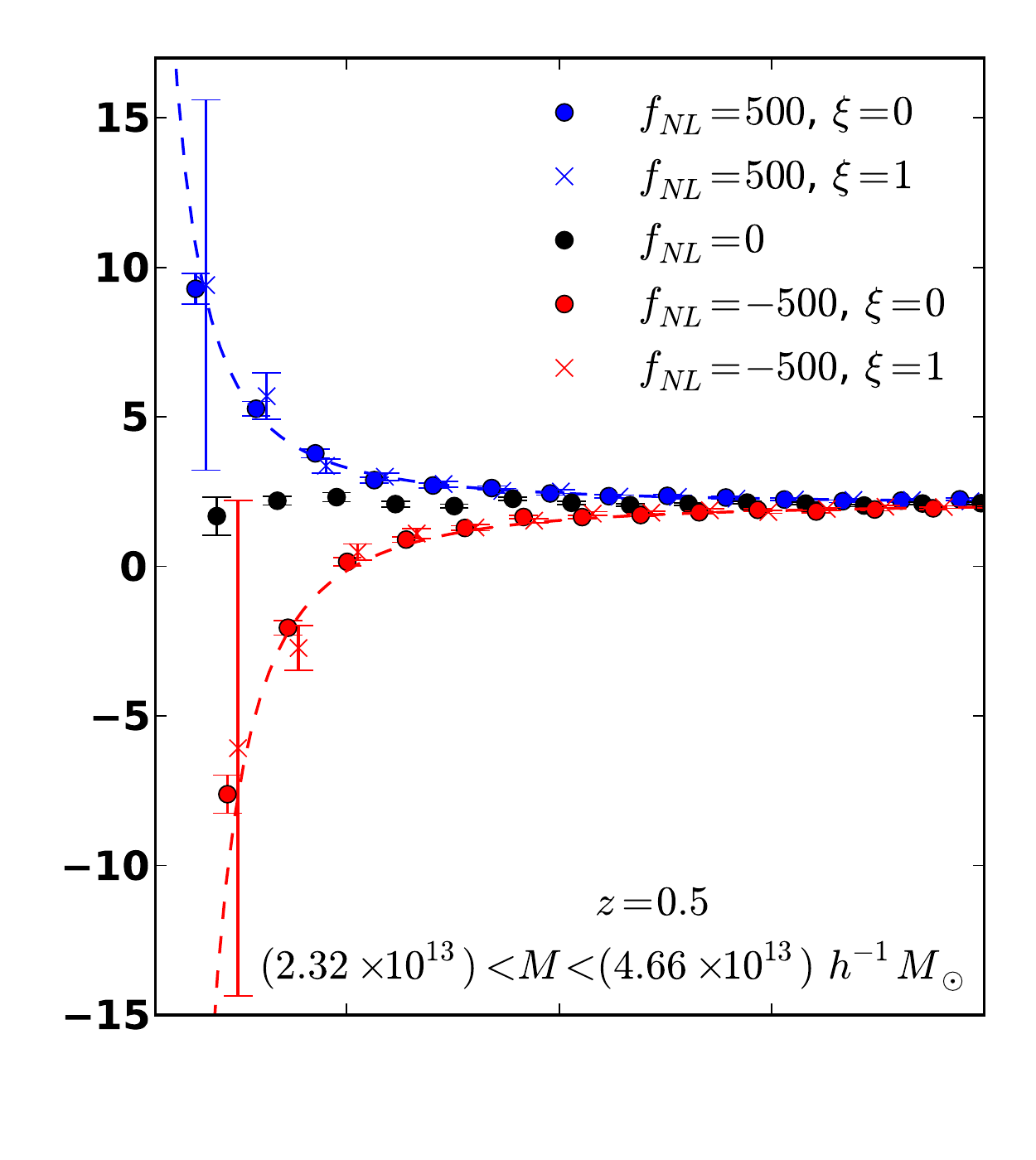}
\includegraphics[width=5.7cm]{xi_bmh_bin1_z05.pdf}
}
\centerline{
\includegraphics[width=5.7cm]{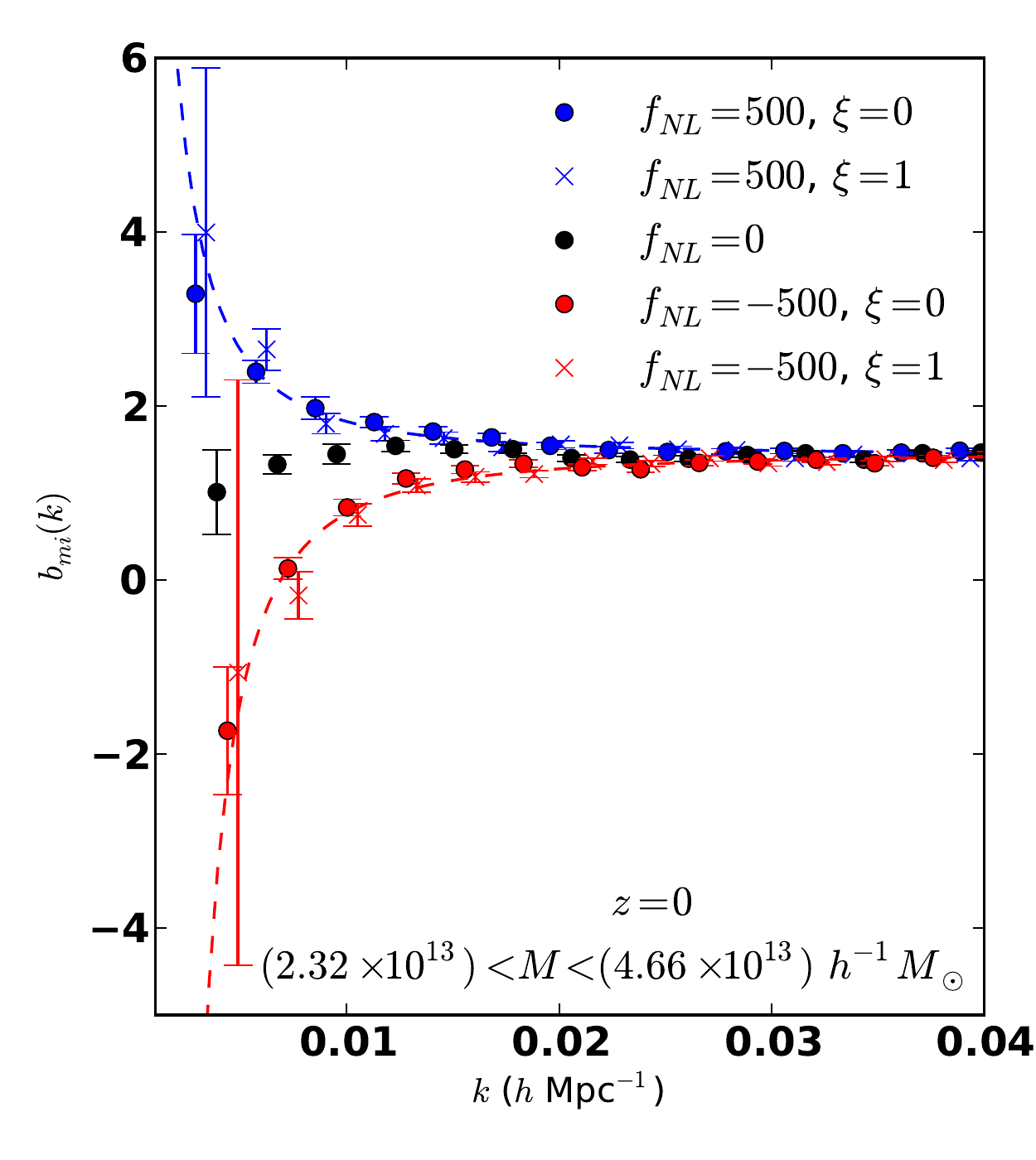}
\includegraphics[width=5.7cm]{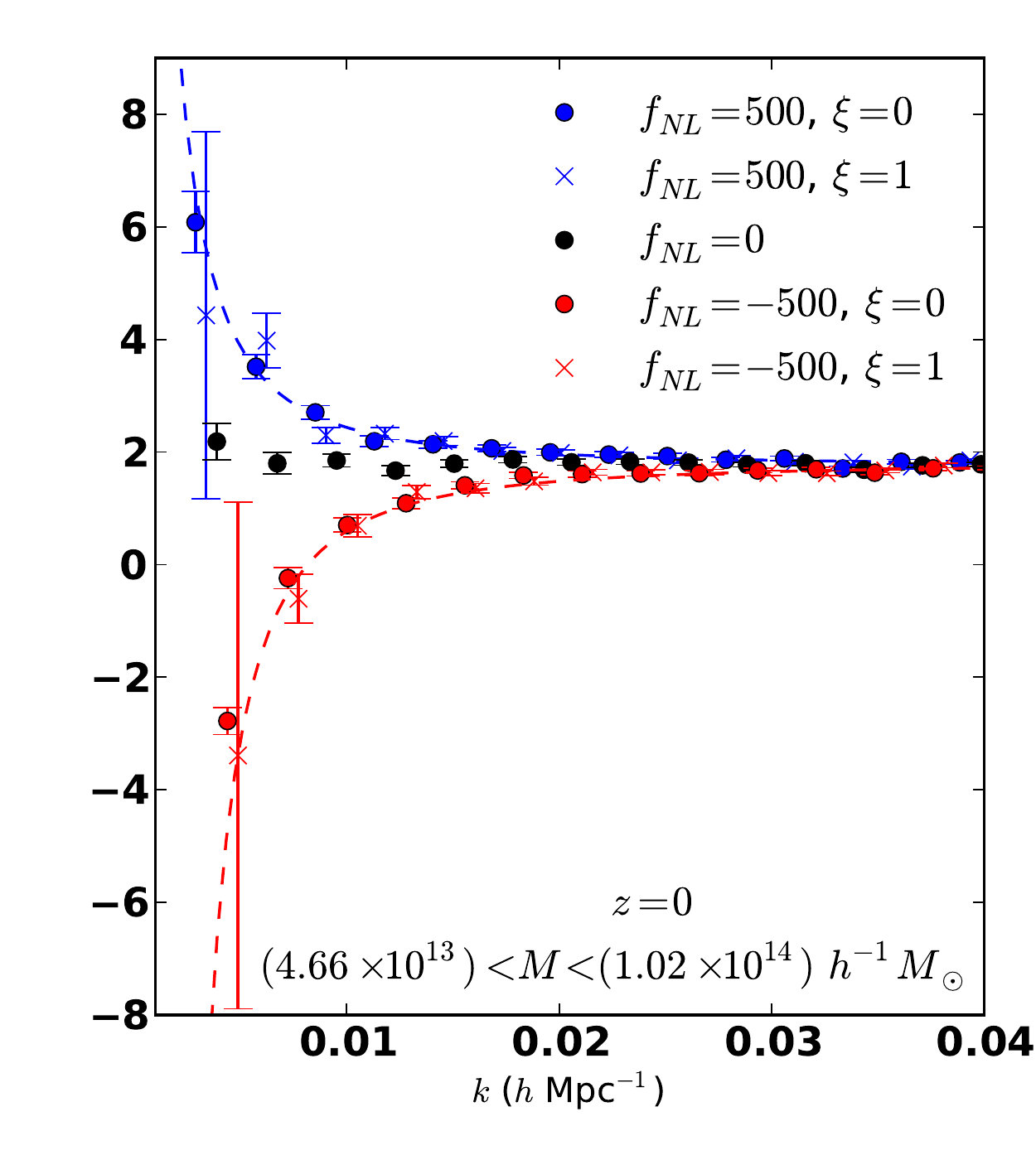}
\includegraphics[width=5.7cm]{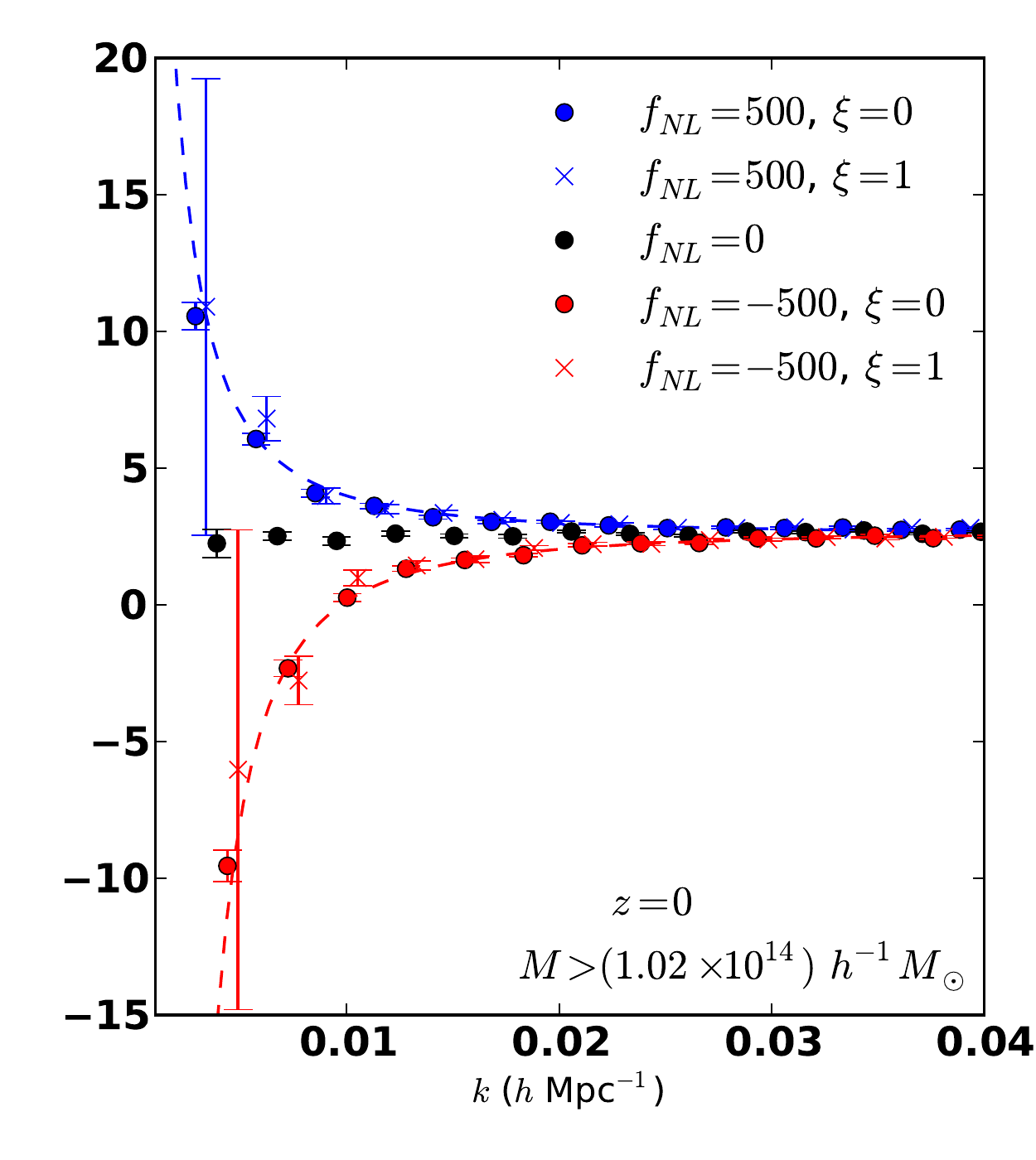}
}
\caption{Halo bias $b_{mi}(k)$ for selected redshifts and halo mass bins, estimated from $N$-body simulations as described in Appendix~\ref{app:estimators}.
The curves are the predicted form in Eq.~(\ref{eq:bias_prediction}), with $b_0$ treated as a free parameter which is fit from data.
}
\label{fig:bias}
\end{figure}

In Fig.~\ref{fig:bias} we show the dependence of $b_{mi}(k)$ on the non-Gaussianity parameters $\fnlcmb$ and $\xi$, for several choices of redshift and mass bin,
on large angular scales ($k \le 0.04$ $h$~Mpc$^{-1}$).

Assigning error bars in Fig.~\ref{fig:bias} is nontrivial.  On large angular scales, variations in $P_{mi}(k)$ and $P_{mm}(k)$ are highly correlated
(since variations in both power spectra are mainly due to sample variance) and therefore mostly cancel when we take the ratio to estimate $b_{mi}(k)$.
In Appendix~\ref{app:estimators} we show in detail how to assign error bars in a way which accounts for this correlation.
The resulting error bars are smaller than sample variance would suggest, and result in very small statistical errors when fitting a given functional form of $b_{mi}(k)$.

Let us separate the issue of whether the predicted form of the bias in Eq.~(\ref{eq:bias_prediction}) agrees with simulation
into three separate questions.

First, we can ask: for Gaussian initial conditions (i.e.~$f_{NL}=0$), is the bias constant on large scales, as predicted by Eq.~(\ref{eq:bias_prediction})?
If we fit for a constant bias for $k \le 0.04$ $h$~Mpc$^{-1}$ in each redshift and mass bin, then we find acceptable $\chi^2$ values for each fit,
indicating that the bias is indeed constant within the statistical errors of the simulations.
This fitting procedure was used to estimate $b_G$ and assign statistical errors in Tab.~\ref{tab:mass_bins}.
Note that the statistical error $\sigma(b_G)$ returned by each fit is typically of order $\approx$0.01, so this $\chi^2$ test shows that the bias
is constant to percent level.

Second, we can ask: if $f_{NL} \ne 0$ but $\xi = 0$, does the predicted form of the non-Gaussian bias in Eq.~(\ref{eq:bias_prediction}) agree with simulation?
We fit for this functional form of $b_{mi}(k)$ over scales $k \le 0.04$ $h$~Mpc$^{-1}$, treating $b_0$ as an independent free parameter for each value of $f_{NL}$, i.e.~we do not 
assume that the constant part of the bias $b_0$ at $\fnlcmb\ne 0$ is equal to the constant Gaussian bias $b_G$ at $\fnlcmb=0$.
In \cite{Desjacques:2008vf}, a slightly different fitting procedure was used: the bias $b_{mi}(k)$ is fit to the functional form\footnote{Note 
that the expression for $\Delta b$ in Eq.~(9) of \cite{Desjacques:2008vf} also includes a term denoted $(b(M) \beta_m(k,\fnlcmb))$ which corresponds to $f_{NL}$ dependence of the matter
power spectrum $P_{mm}(k)$.  We do not include this term because we define the bias to be $b_{mi}(k,f_{NL}) = P_{mi}(k,f_{NL}) / P_{mm}(k,f_{NL})$ with an
$f_{NL}$-dependent denominator.  Such a term would be needed if the bias were defined as $(P_{mi}(k,f_{NL})/P_{mm}(k,0))$ or as $(P_{mi}(k,f_{NL})/P_{\rm lin}(k))$.}
\be
b_{mi}(k) = b_G + \frac{2 \delta_c}{\alpha(k,z)} \fnlcmb (b_G-1) + (\Delta b_I)  \label{eq:desjacques_fit}
\ee
where $(\Delta b_I)$ is derived by replacing $\bar{n}_h$ in Eq. (\ref{eq:bGdef}) with a non-Gaussian mass function and $b_G$ is the Gaussian (i.e.~$\fnlcmb=0$) bias.
This functional form is not precisely equivalent to Eq.~(\ref{eq:bias_prediction}): the two differ at $\bigoh(\fnlcmb^2)$ by the term 
$(2\delta_c \fnlcmb/\alpha(k,z)) (\Delta b_I)$. We have not investigated whether one of the functional forms is a better fit than the other, but we anticipate that an $\bigoh(\fnlcmb^2)$  difference will be negligible for values of $\fnlcmb$ which are observationally relevant. However, we do find that including an $\bigoh(\fnlcmb)$ term which is constant in $k$, either via the last term in Eq.~(\ref{eq:desjacques_fit}) or by allowing $b_0$ to differ from $b_G$ in Eq.~(\ref{eq:bias_prediction}), is needed to obtain a good fit to simulation. The magnitude of this scale-independent $f_{NL}$ correction we find is in qualitative agreement with what one would get using the non-Gaussian mass-function of \cite{LoVerde:2007ri} in Eq. (\ref{eq:bGdef}).

We find that for some choices of redshift and halo mass bin, the fits return bad $\chi^2$ values (Tab.~\ref{tab:bias_chisquare}), i.e.~we find statistically significant disagreement
between the simulations and the predicted form of the bias in Eq.~(\ref{eq:bias_prediction}).
Detailed inspection of the bad fits shows that, in all cases, the prediction tends to overestimate the magnitude of the non-Gaussian bias on large scales, but
the discrepancy between simulation and prediction is only $\approx$10\% of the total non-Gaussian bias in the worst case.
The theoretical assumptions made in deriving the non-Gaussian bias, i.e.~a universal mass function and the peak-background split relation between halo bias and the mass function,
also fail at roughly this level \cite{Robertson:2008jr,Tinker:2010my}, so a $\approx$10\% discrepancy is not at all surprising.

\begin{table}
\begin{center}
\begin{tabular}{|c|c||c|c||c|c|}
\hline &                               & \multicolumn{2}{c}{$f_{NL}=500$} & \multicolumn{2}{|c|}{$f_{NL}=-500$} \\ \cline{3-6}
& Mass range ($h^{-1} M_\odot$) & $b_0$ & $\chi^2/N_{\rm dof}$ & $b_0$ & $\chi^2/N_{\rm dof}$ \\ \hline\hline
$z=2$ & $M > 1.15 \times 10^{13}$ & $4.228 \pm 0.020$ & 61.6/13 & $6.560 \pm 0.083$ & 70.7/13 \\
\hline $z=1$ & $1.15 \times 10^{13} < M < 2.32\times 10^{13}$ & $2.238 \pm 0.008$ & 29.6/13 & $2.878 \pm 0.023$ & 65.9/13 \\
      & $M > 2.32 \times 10^{13}$ & $3.012 \pm 0.011$ & 42.9/13 & $4.141 \pm 0.033$ & 46.4/13 \\
\hline $z=0.5$ & $1.15 \times 10^{13} < M < 2.32\times 10^{13}$ & $1.623 \pm 0.007$ & 28.3/13 & $1.867 \pm 0.015$ & 40.4/13 \\
        & $2.32 \times 10^{13} < M < 4.66\times 10^{13}$ & $1.940 \pm 0.010$ & 11.8/13 & $2.346 \pm 0.022$ & 25.4/13 \\
        & $M > 4.66 \times 10^{13}$ & $2.666 \pm 0.010$ & 36.4/13 & $3.448 \pm 0.026$ & 31.0/13 \\
\hline $z=0$ & $1.15 \times 10^{13} < M < 2.32\times 10^{13}$ & $1.193 \pm 0.006$ & 27.6/13 & $1.229 \pm 0.010$ & 17.3/13 \\
      & $2.32 \times 10^{13} < M < 4.66\times 10^{13}$ & $1.386 \pm 0.008$ & 7.7/13 & $1.524 \pm 0.014$ & 23.2/13 \\
      & $4.66 \times 10^{13} < M < 1.02\times 10^{14}$ & $1.675 \pm 0.010$ & 14.9/13 & $1.907 \pm 0.018$ & 30.5/13 \\
      & $M > 1.02\times 10^{14}$ & $2.403 \pm 0.010$ & 23.8/13 & $2.946 \pm 0.022$ & 20.1/13 \\
\hline
\end{tabular}
\end{center}
\caption{Best-fit values of $b_0$, and $\chi^2$ values for the fit, when fitting the predicted form of the non-Gaussian bias
in Eq.~(\ref{eq:bias_prediction}) to estimates of the bias $b_{mi}(k)$ from simulation, with error bars assigned as described
in Appendix~\ref{app:estimators}.  A few of the fits return bad $\chi^2$ values; in these cases we find that Eq.~(\ref{eq:bias_prediction})
overpredicts the non-Gaussian bias by $\approx$10\%.}
\label{tab:bias_chisquare}
\end{table}

Third, we can ask whether the bias $b_{mi}(k)$ is independent of $\xi$ for fixed $\fnlcmb$, as predicted by Eq.~(\ref{eq:bias_prediction}).
To test this, we let $b(k)$ and $b'(k)$ denote estimates of the bias from two $N$-body simulations with $\xi=0$ and $\xi=1$ (and the same value of $\fnlcmb$).
We define a $\chi^2$ statistic by summing over $k$-bins:
\be
\chi^2 = \sum_k \frac{(b(k) - b'(k))^2}{\Var(\Delta b(k)) + \Var(\Delta b'(k))}
\ee
For almost all redshifts and mass bins, we find acceptable $\chi^2$ values, indicating that the bias estimates from the two simulations
are consistent within statistical errors.
(There is one exception: we find an anomalous $\chi^2$ for $\fnlcmb=-500$ and $z=2$, but inspection of the bad fit shows that the bias only
differs by $\approx 10\%$ between $\xi=0$ and $\xi=1$.  The bias is larger in the $\xi=1$ case, at low $k$.)

Our conclusion in this section is that the prediction for the non-Gaussian halo bias in Eq.~(\ref{eq:bias_prediction}) is an impressive fit to the simulations
across a wide range of curvaton model parameters.
Although we do detect statistically significant deviations from the prediction at the $\approx 10\%$ level, this is typical for results based on general
arguments such as the peak-background split.

\section{Halo stochasticity}
\label{ssec:halo_stochasticity}

For halo mass bins $i,j$, we define the stochasticity parameter
\be
r_{ij}(k) = \frac{P_{ij}(k) - \delta_{ij}/n_i}{P_{mm}(k)} - \frac{P_{mi}(k) P_{mj}(k)}{P_{mm}(k)^2}\,.  \label{eq:rijdef}
\ee
If the halos are perfectly non-stochastic tracers of the dark matter (or more precisely, if the only source of stochasticity is shot noise)
then both the diagonal (i.e. $i=j$) and non-diagonal (i.e. $i\ne j$) components of $r_{ij}(k)$ will be zero.
As discussed in \S\ref{ssec:halo_model}, the halo model predicts that the leading contribution to $r_{ij}(k)$ on large scales arises from 1-halo
terms in the matter-halo power spectra.  If $\xi > 0$, then we expect large $r_{ij}$ since the non-Gaussian part of the bias will
be stochastic.

\begin{figure}[!ht]
\centerline{
\includegraphics[width=5.7cm]{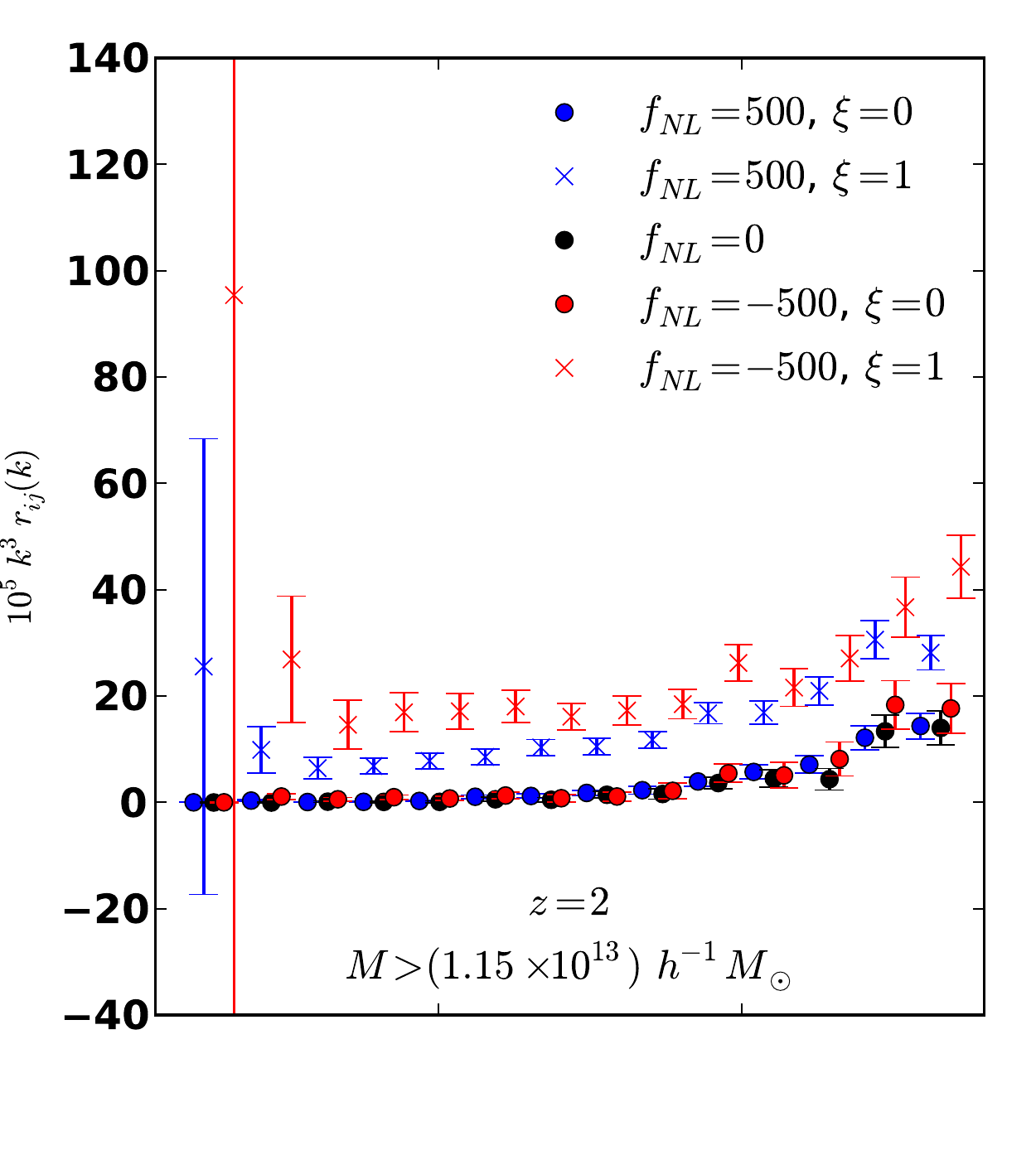}
\includegraphics[width=5.7cm]{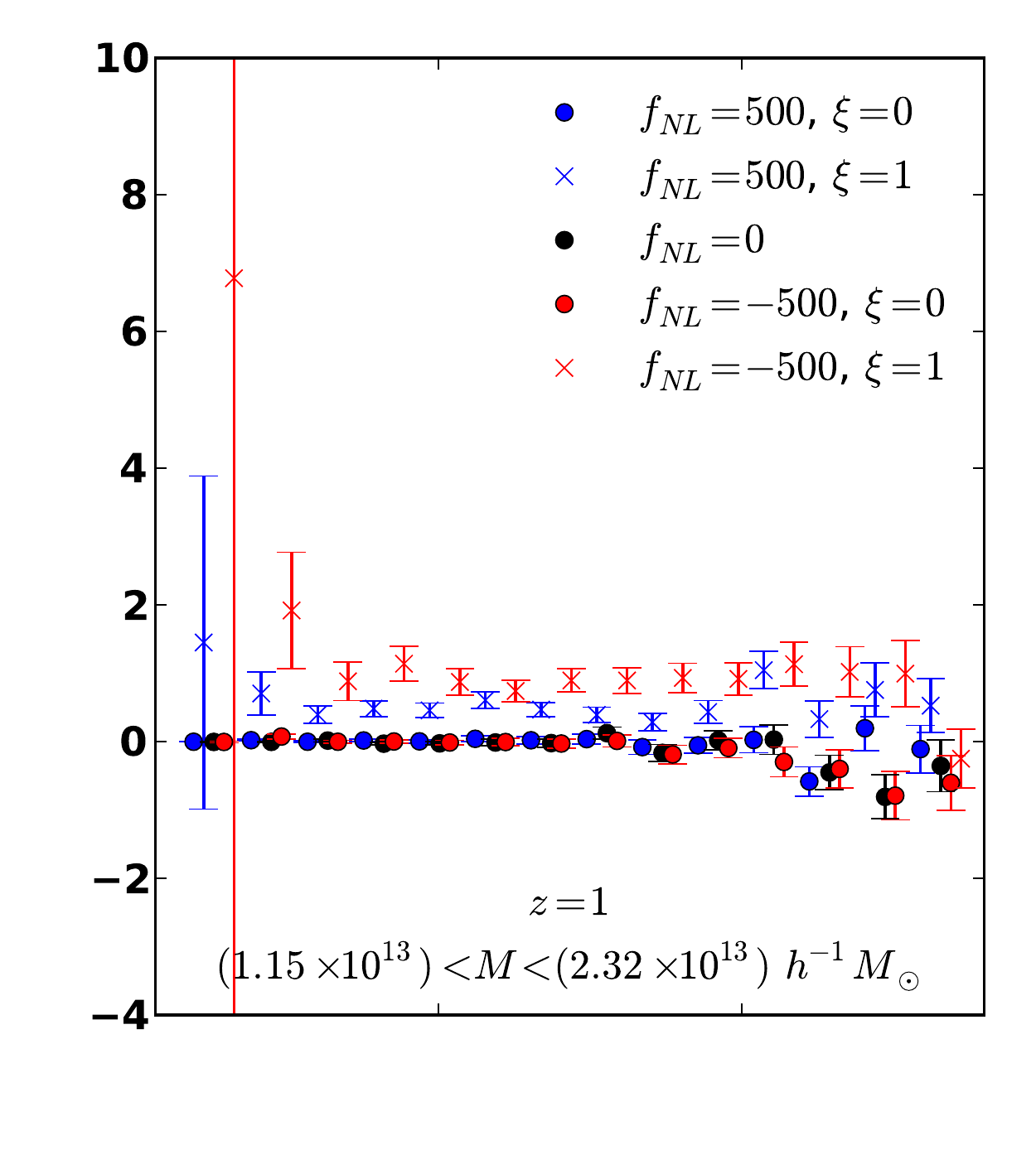}
\includegraphics[width=5.7cm]{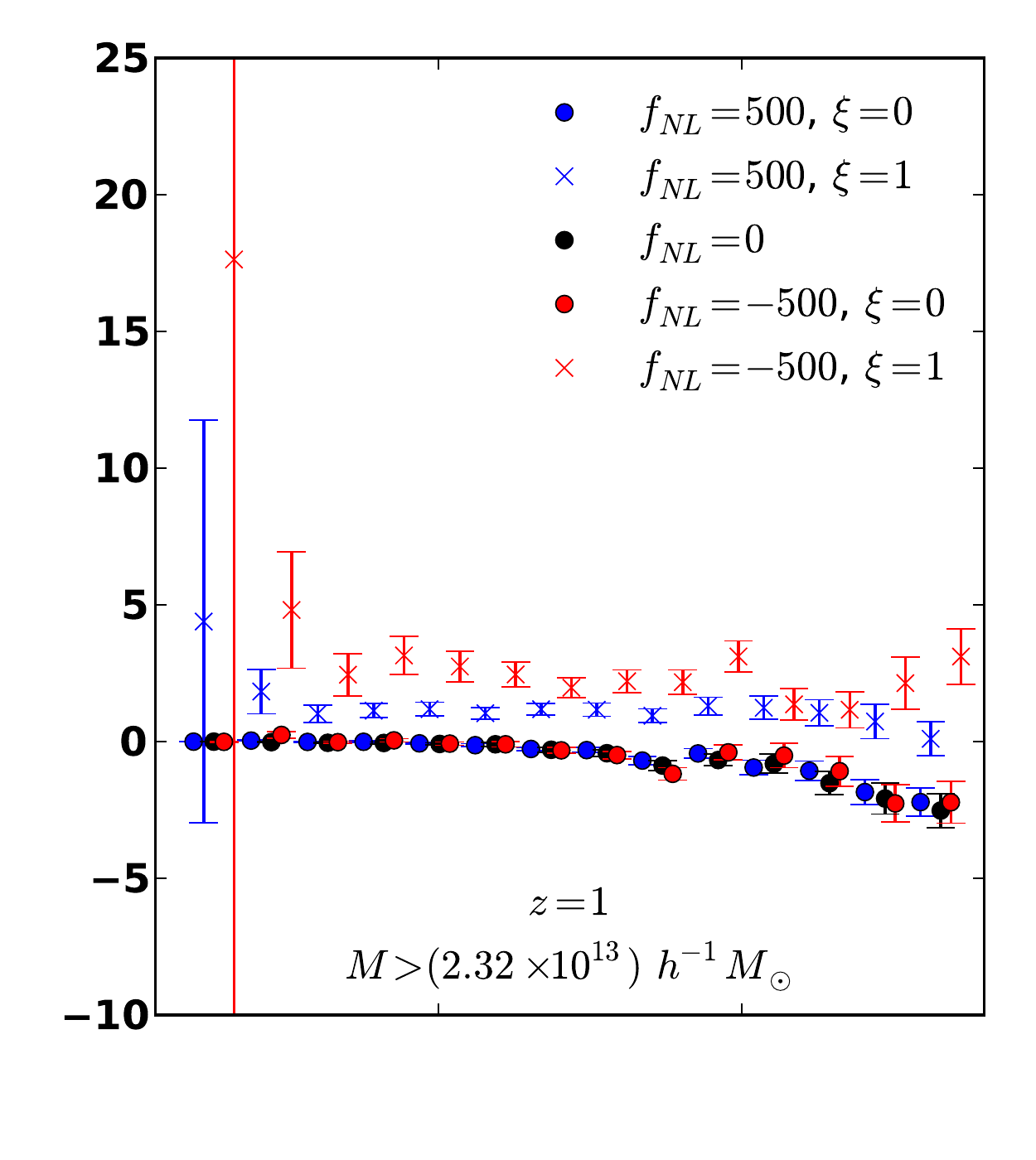}
}
\centerline{
\includegraphics[width=5.7cm]{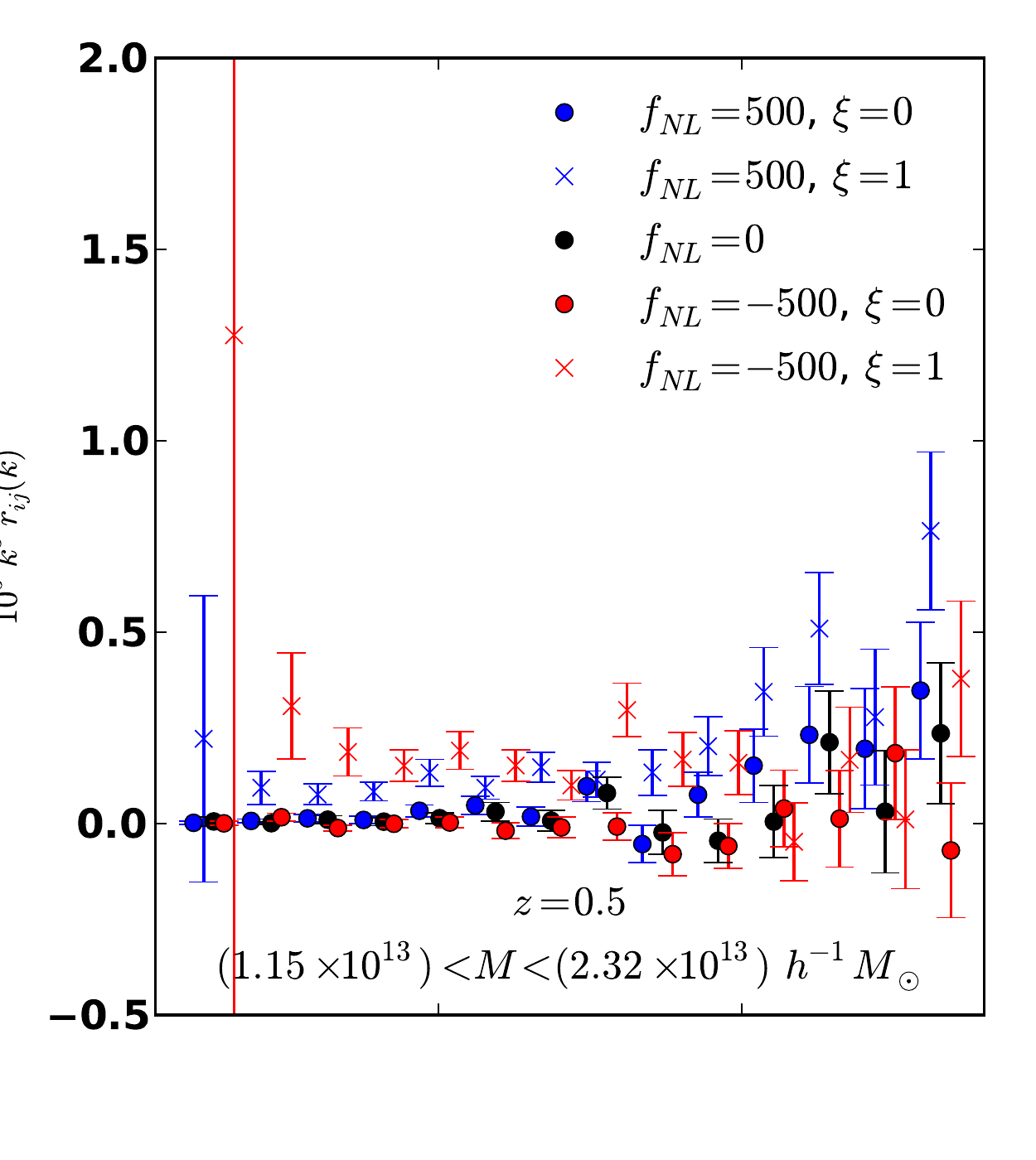}
\includegraphics[width=5.7cm]{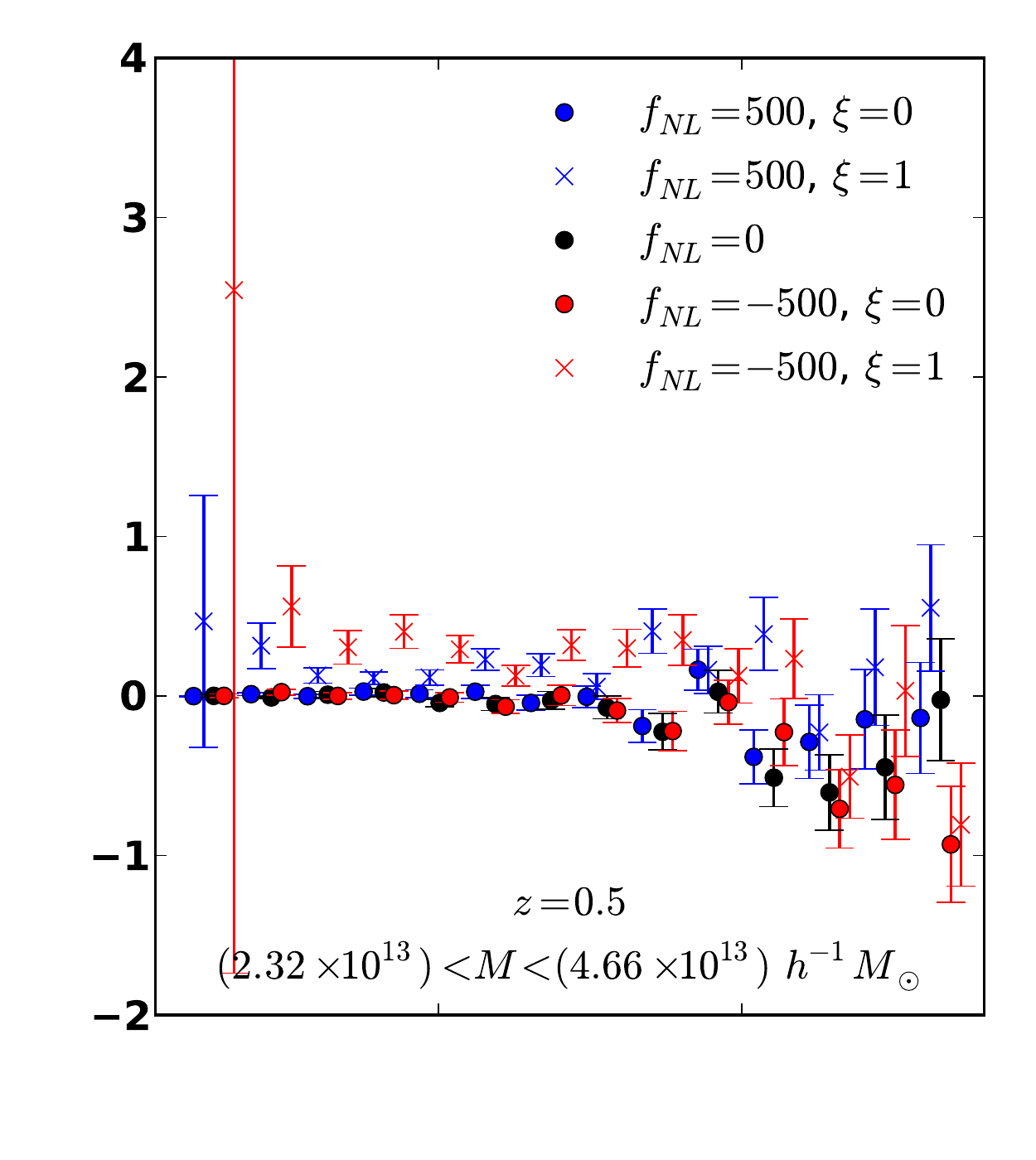}
\includegraphics[width=5.7cm]{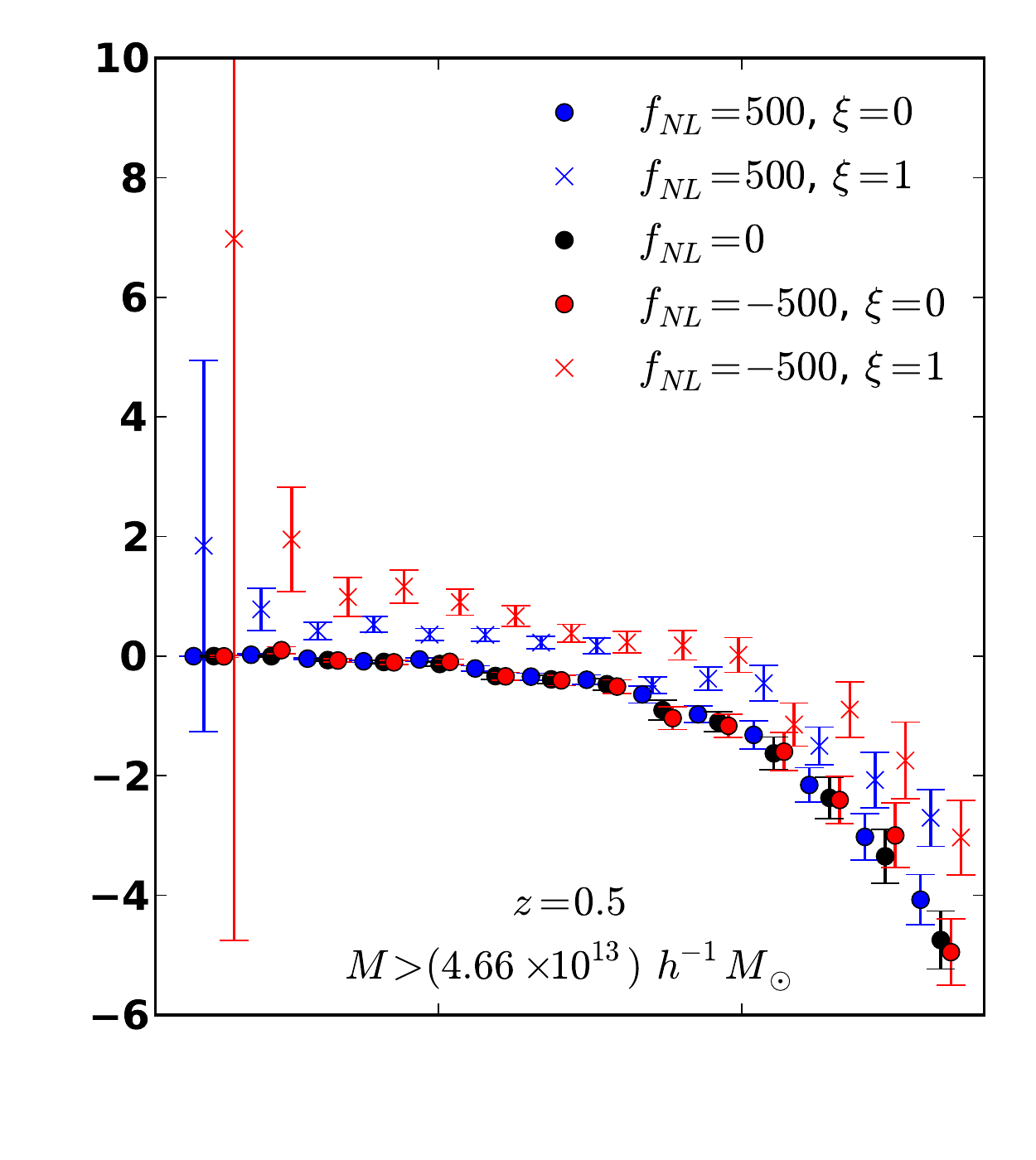}
}
\centerline{
\includegraphics[width=5.7cm]{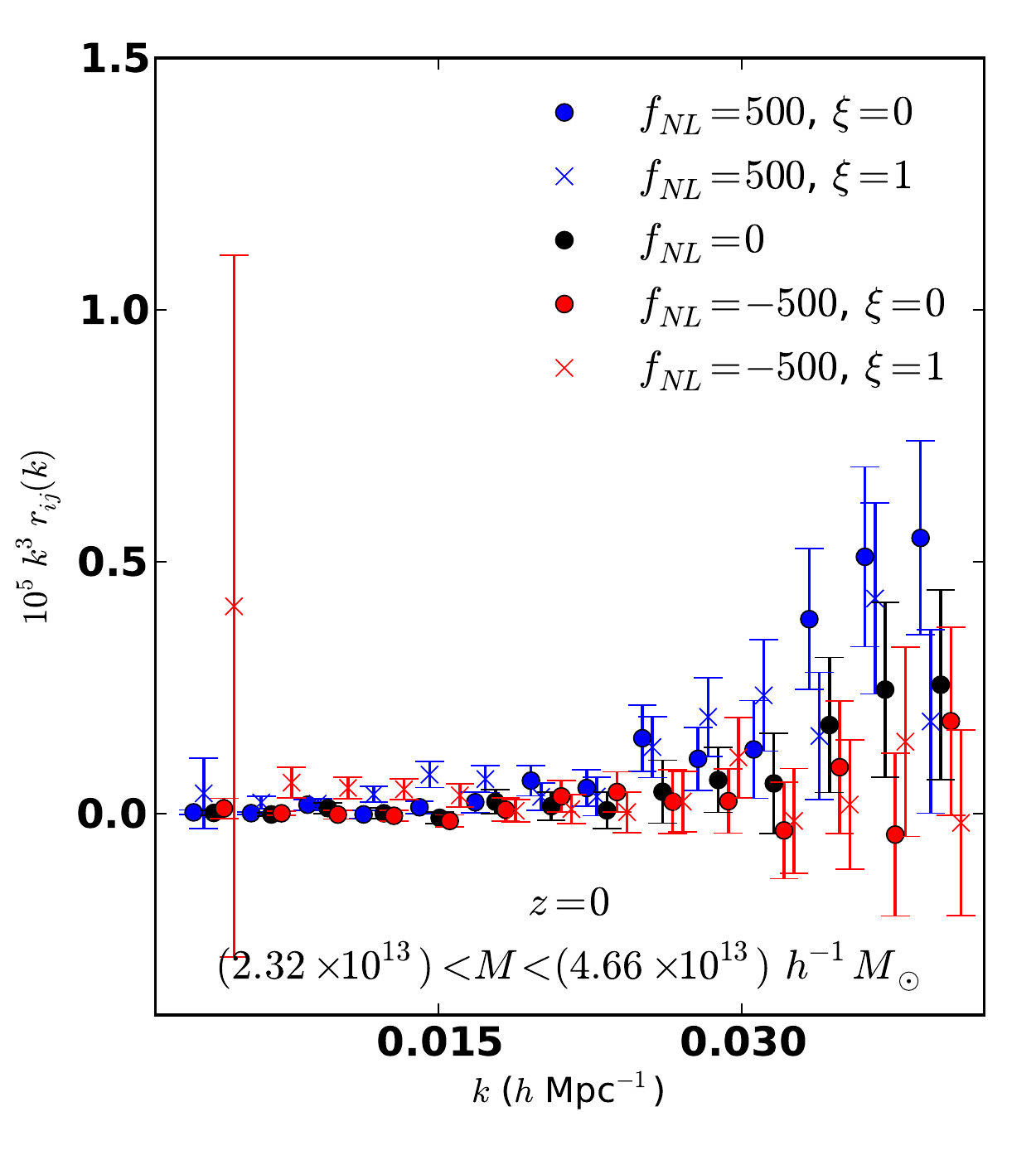}
\includegraphics[width=5.7cm]{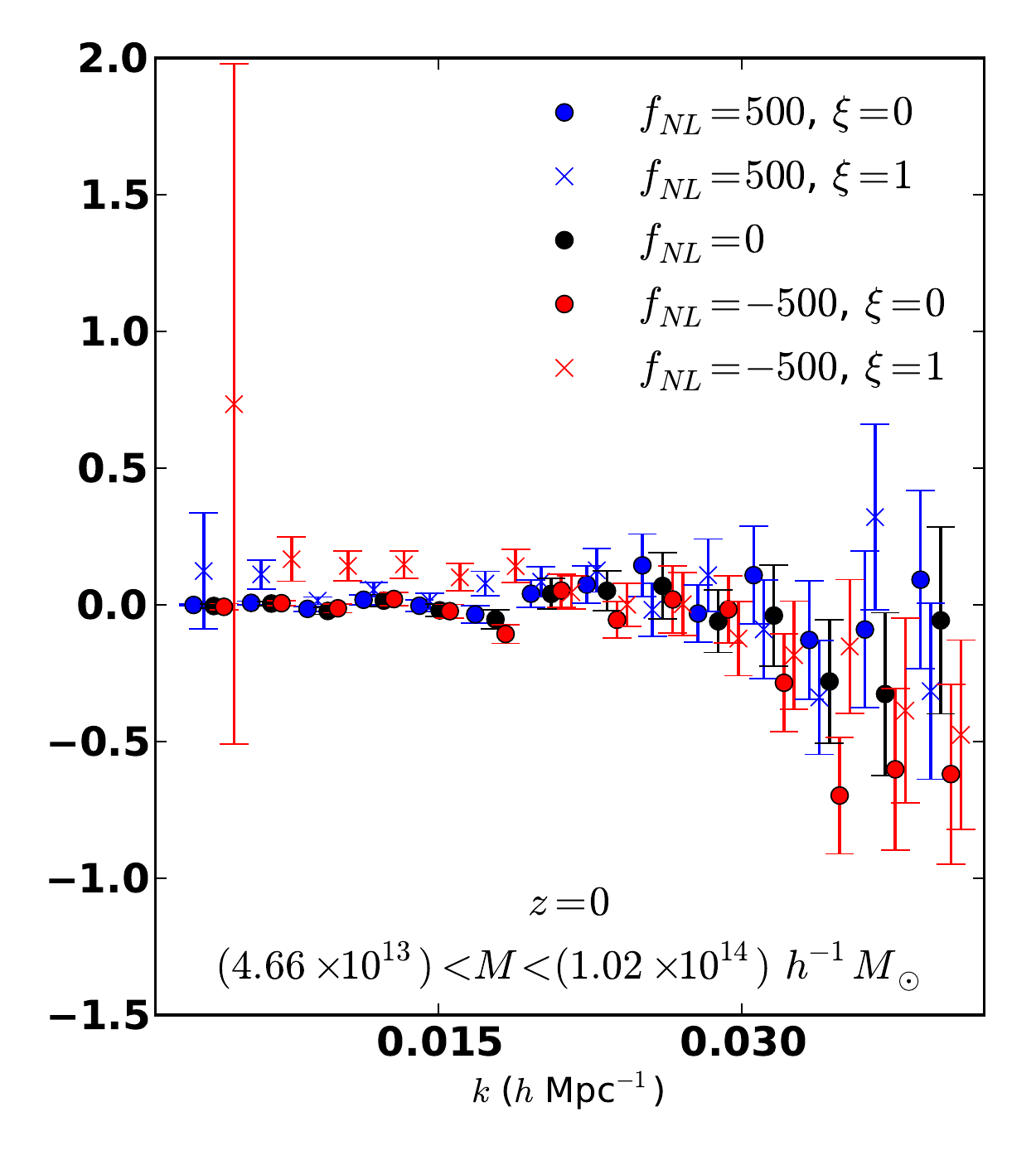}
\includegraphics[width=5.7cm]{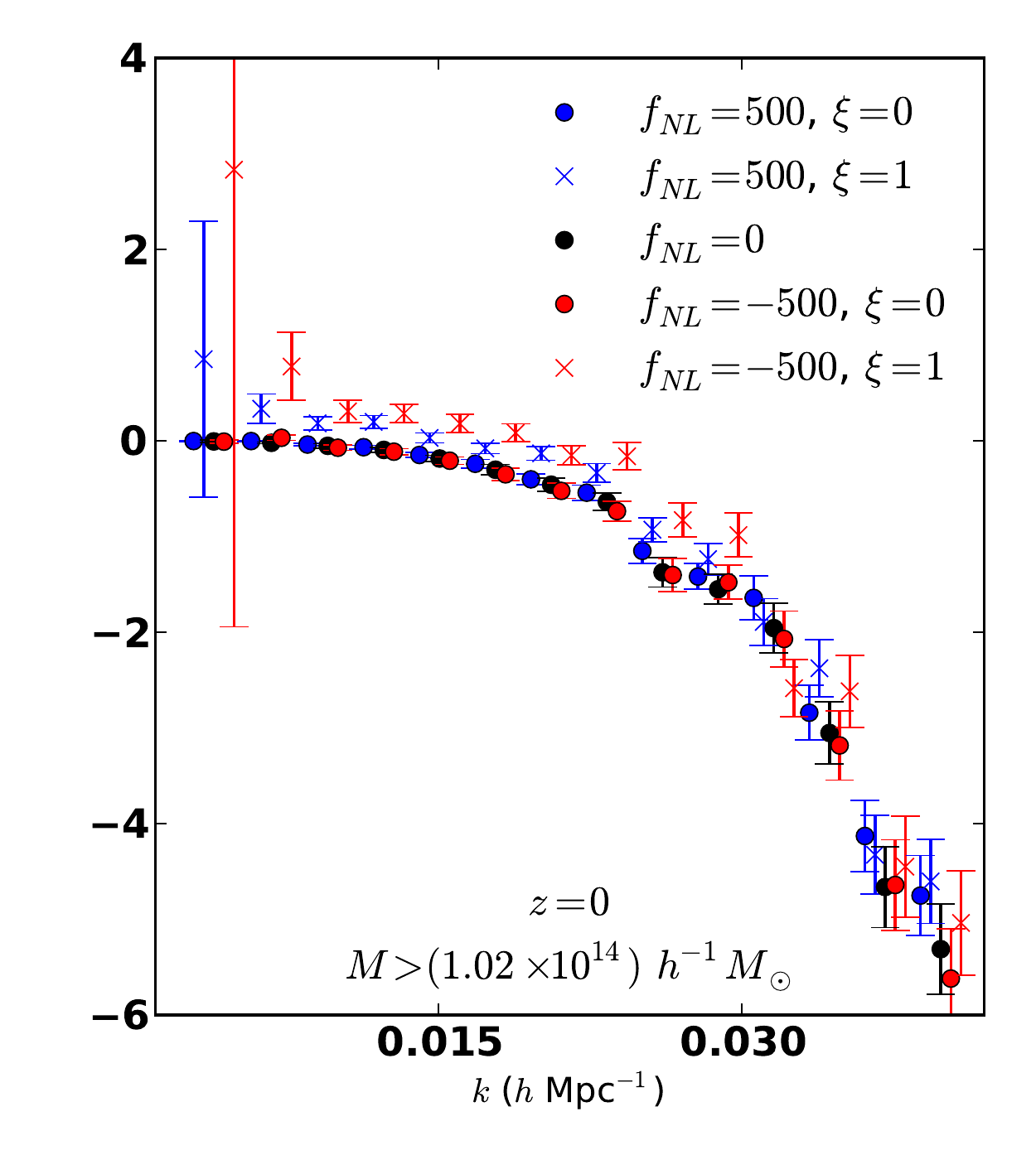}
}
\caption{Diagonal components $r_{ii}$ of the stochasticity statistic defined in Eq.~(\ref{eq:rijdef}), estimated from $N$-body simulations as described in
Appendix~\ref{app:estimators}, for varying choices of redshift, halo mass range, and curvaton model parameters ($\fnlcmb$, $\xi$).}
\label{fig:stochasticity}
\end{figure}

In Fig.~\ref{fig:stochasticity}, we show estimates of the diagonal components $r_{ii}(k)$ from simulation, for several choices of redshift,
mass bin, and model parameters ($\fnlcmb$, $\xi$).
Similarly to the case of halo bias from the previous section, assigning error bars is nontrivial, since estimates of $P_{mm}$, $P_{mi}$, and $P_{ij}$ are all highly
correlated on large scales.
Our procedure for estimating the stochasticity and assigning error bars is given in Appendix~\ref{app:estimators} and results in error bars
which are much smaller than sample variance would suggest.

In this section, we will compare the stochasticity estimated from simulations to the predicted form
\ba
r_{ij}(k) &=& \Big[ (b^{(i)}_G + b^{(i)}_{NG}(k))(b^{(j)}_G + b^{(j)}_{NG}(k)) + \xi^2 b^{(i)}_{NG}(k) b^{(j)}_{NG}(k) \Big] \frac{P_{\rm lin}(k)}{P_{\rm lin}(k) + P_{mm}^{1H}} \nn \\
  && \hspace{0.5cm} - \, \frac{\Big[ (b^{(i)}_G + b^{(i)}_{NG}(k)) P_{\rm lin}(k) + f_i/n_i \Big] \Big[ (b^{(j)}_G + b^{(j)}_{NG}(k)) P_{\rm lin}(k) + f_j/n_j \Big]}{(P_{\rm lin}(k)+P_{mm}^{1H})^2} 
\label{eq:stochasticity_prediction}
\ea
which follows from the halo model calculations in \S\ref{ssec:halo_model}.
As in the previous section, we will separate the issue of whether this prediction agrees with simulation into three separate questions.

\begin{figure}
\centerline{\includegraphics[width=12cm]{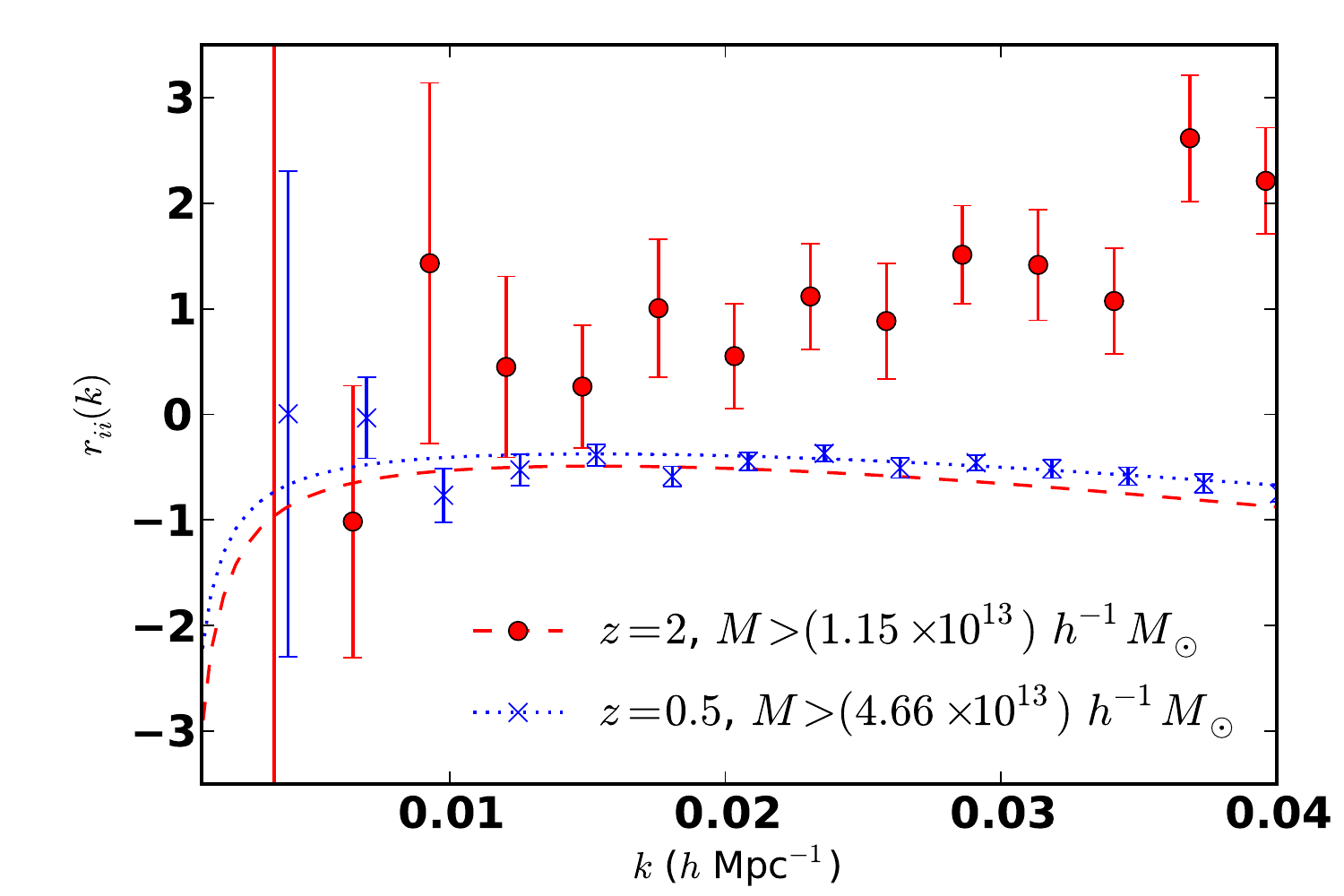}}
\caption{Stochasticity parameter $r_{ii}$ estimated from {\em Gaussian} simulations (error bars), with halo model
prediction shown for comparison (curves).  In general, we do not find that the halo model accurately predicts $r_{ii}$.
An example of a redshift and mass bin where the halo prediction disagrees with simulation ($z=2$ and $M > 1.15 \times 10^{13}$ $h^{-1}M_\odot$) 
and an example where the two agree ($z=0.5$ and $M > 4.66 \times 10^{13}$ $h^{-1}M_\odot$) are shown.
}
\label{fig:rii_gaussian}
\end{figure}

First, does the prediction in Eq.~(\ref{eq:stochasticity_prediction}) agree with simulations in the Gaussian case ($f_{NL}=0$)?
Surprisingly, we do not even find agreement at a qualitative level.
Although there are a few choices of redshift and mass bin where the halo model prediction roughly fits the data, there are more
cases where there is no resemblance (example good and bad fits are shown in Fig.~\ref{fig:rii_gaussian}).\footnote{A
recent paper \cite{Hamaus:2010im} also compared stochasticity predictions in the halo model with $N$-body simulations in the Gaussian
case.  It was found that the smallest eigenvalue of the stochasticity matrix $r_{ij}$ is predicted accurately by
the halo model. (The smallest eigenvalue is particularly relevant since it corresponds to a halo mass weighting with reduced
shot noise \cite{Seljak:2009af}.)  However, it can also be seen (Fig.~12 of \cite{Hamaus:2010im}) that the halo model does not
accuractely predict the largest eigenvalue, which implies that some matrix elements $r_{ij}$ are not accurately predicted.
In this paper, we have concentrated on the diagonal $r_{ii}$ and do not generally find good agreement with the halo model.}

In this paper, we have not attempted to propose a general model for $r_{ii}$ in the Gaussian case.
Therefore, when we compare the prediction for $r_{ij}$ to simulation in the non-Gaussian case,
our approach is to estimate the {\em change} in stochasticity $\Delta r_{ij}$ between non-Gaussian and Gaussian initial conditions, as a function
of $(\fnlcmb,\xi)$, and compare with the prediction for $\Delta r_{ij}$ obtained from Eq.~(\ref{eq:stochasticity_prediction}).

The second question we can ask is, how does the stochasticity $r_{ii}$ depend on $\fnlcmb$, in the case $\xi=0$?
In this case, we find that Eq.~(\ref{eq:stochasticity_prediction}) predicits an $\fnlcmb$ dependence which is small compared to the
statistical errors of our simulations.
To test this prediction, we define a $\chi^2$ statistic by:
\be
\chi^2 = \sum_k \frac{(\hat r_{ii}(k) - \hat r'_{ii}(k))^2}{(\Delta r_{ii}(k))^2 + (\Delta r'_{ii}(k))^2}
\ee
where $\hat r_{ii}(k)$ and $\hat r'_{ii}(k)$ denote stochasticity estimates from $N$-body simulations with $f_{NL} = 0$ and $f_{NL} \ne 0$
respectively (taking $\xi=0$ in the non-Gaussian simulation).
For $f_{NL} = \pm 500$ and all redshifts and mass bins considered in this paper, we find good $\chi^2$ values, i.e.~no
statistically significant dependence of the stochasticity $r_{ii}$ on $\fnlcmb$ (provided $\xi=0$).
This agrees with the halo model prediction~(\ref{eq:stochasticity_prediction}), even though the halo model does not correctly predict
the actual value of $r_{ii}$ as previously remarked.

Third, we can ask, how does the stochasticity depend on $\xi$?
Since we do not have a model for the Gaussian stochasticity, we ask whether the quantity
\be
\Delta r_{ii} = r_{ii}(k,\fnlcmb,\xi) - r_{ii}(k,\fnlcmb=0)
\ee
i.e.~the excess stochasticity over Gaussian, is correctly modeled by the peak-background split prediction:
\be
\Delta r_{ii} \approx \xi^2 b_{NG}(k)^2 = \left( \xi \fnlcmb \frac{ 2 \delta_c (b_0 - 1) }{\alpha(k,z)} \right)^2  \label{eq:delta_rii_prediction}
\ee
(This is actually an approximation to the prediction obtained by differencing Eq.~(\ref{eq:stochasticity_prediction})
between values of $(\fnlcmb,\xi)$, but we find that this approximation is within statistical errors of the simulation,
so it is a convenient simplification.)

\begin{figure}[!h]
\centerline{\includegraphics[width=12cm]{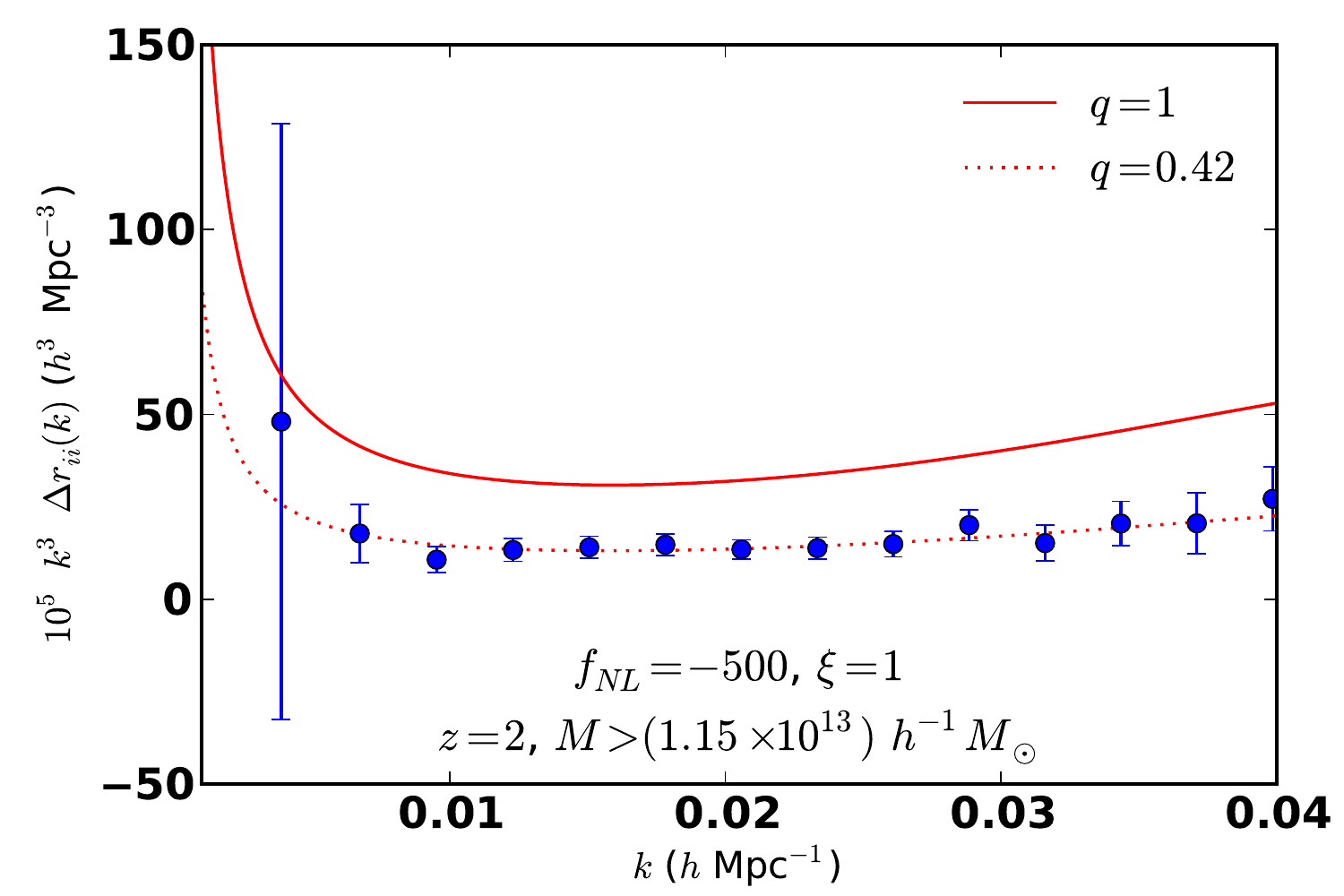}}
\caption{Change in stochasticity parameter $\Delta r_{ii} = r_{ii}(k,\fnlcmb,\xi) - r_{ii}(k,\fnlcmb=0)$ between the
curvaton model with $(f_{NL},\xi)=(500,1)$ and the Gaussian case, estimated from $N$-body simulations.
The peak-background split (solid curve) overpredicts the level of stochasticity, but excellent agreement with simulation is obtained
by scaling the prediction by $q=0.42$ (dotted).
When the parameters ($f_{NL}$,$\xi$,$z$) and the halo mass range are varied,
we find that scaling the peak-background split prediction always provides a good fit, but the value
of $q$ varies, as shown in Tab.~\ref{tab:q}.
}
\label{fig:delta_rii_example_fit}
\end{figure}

We find that the prediction in Eq.~(\ref{eq:delta_rii_prediction}) systematically overestimates $(\Delta r_{ii})$.
Empirically, we find that if we scale the prediction for $(\Delta r_{ii})$ by a multiplicative constant $q$, then the modified 
prediction
\be
\Delta r_{ii} = q \left( \xi \fnlcmb \frac{ 2 \delta_c (b_0 - 1) }{\alpha(k,z)} \right)^2  \label{eq:q_fudge}
\ee
is an excellent fit to the simulations (i.e.~when $q$ is treated as a free parameter, all fits have good $\chi^2$ values), 
for all choices of redshfit, halo mass bin, and $\fnlcmb$.
In Fig.~\ref{fig:delta_rii_example_fit}, we show an example fit; it is seen that the peak-background split overpredicts the
level of stochasiticity, but an excellent fit is obtained by simply scaling the peak-background split prediction.
In Tab.~\ref{tab:q}, we tabulate values of $q$ obtained by fitting Eq.~(\ref{eq:q_fudge}) to the simulations, together with statistical
errors from the fits.

One general trend evident in this table is that $q$ is an increasing function of $\fnlcmb$.
This simply means that we are considering large enough values of $\fnlcmb$ that we are sensitive to $\bigoh(\fnlcmb^3)$ terms,
if we think of $(\Delta r_{ii})$ as a power series in $\fnlcmb$.
However, it is clear from Tab.~\ref{tab:q} that if we interpolate to $\fnlcmb\rightarrow 0$, most values of $q$ are still $< 1$.
This indicates that, even to leading order $\bigoh(\fnlcmb^2)$, we are seeing $\approx 30\%$ disagreement between the 
prediction~(\ref{eq:delta_rii_prediction}) and simulation.
We have not attempted to study $\bigoh(\fnlcmb^3)$ terms in detail, or compare them between theory and simulation, since we are 
already seeing disagreement at leading order $\bigoh(\fnlcmb^2)$.

Since the typical value is $q \approx 0.7$, our interpretation is that the peak-background split generally overpredicts non-Gaussian
stochasticity by $\approx$30\%.
Although some discrepancy is expected (e.g.~in the previous section we found $\lsim 10$\% discrepancies in predictions for halo bias),
this level of disagreement is somewhat uncomfortable and should probably be incorporated when constraining $\xi$ from observations.
It is not clear how to interpret this discrepancy theoretically, or whether it is related to the discrepancy that we found previously
in the Gaussian case (Fig.~\ref{fig:rii_gaussian}).
Modeling stochasticity using a semianalytic framework such as the peak-background split or halo model appears to be more difficult
than modeling halo bias.

\begin{table}[!h]
\begin{center}
\begin{tabular}{|c|c|c|c|c|c|}
\hline & Mass range ($h^{-1} M_\odot$) & $\fnlcmb=500$ & $\fnlcmb=250$ & $\fnlcmb=-250$ & $\fnlcmb=-500$ \\ \hline\hline
$z=2$ & $M > 1.15 \times 10^{13}$  &  $0.98 \pm 0.07$  &  $0.88 \pm 0.08$  &  $0.62 \pm 0.06$  &  $0.42 \pm 0.03$ \\
\hline $z=1$ & $1.15 \times 10^{13} < M < 2.32\times 10^{13}$  &  $0.79 \pm 0.09$  &  $0.83 \pm 0.12$  &  $0.67 \pm 0.09$  &  $0.46 \pm 0.04$ \\
      & $M > 2.32 \times 10^{13}$  &  $0.83 \pm 0.07$  &  $0.70 \pm 0.08$  &  $0.66 \pm 0.07$  &  $0.51 \pm 0.04$ \\
\hline $z=0.5$ & $1.15 \times 10^{13} < M < 2.32\times 10^{13}$  &  $1.01 \pm 0.18$  &  $0.92 \pm 0.29$  &  $0.45 \pm 0.19$  &  $0.57 \pm 0.10$ \\
        & $2.32 \times 10^{13} < M < 4.66\times 10^{13}$  &  $0.80 \pm 0.15$  &  $0.58 \pm 0.22$  &  $0.73 \pm 0.19$  &  $0.48 \pm 0.08$ \\
        & $M > 4.66 \times 10^{13}$  &  $0.81 \pm 0.09$  &  $0.79 \pm 0.12$  &  $0.80 \pm 0.10$  &  $0.51 \pm 0.05$ \\
\hline $z=0$ & $1.15 \times 10^{13} < M < 2.32\times 10^{13}$  &  $1.37 \pm 0.80$  &  $1.06 \pm 1.12$  &  $1.00 \pm 1.41$  &  $0.90 \pm 0.51$ \\
      & $2.32 \times 10^{13} < M < 4.66\times 10^{13}$  &  $1.35 \pm 0.44$  &  $1.57 \pm 0.77$  &  $0.82 \pm 0.59$  &  $0.58 \pm 0.25$ \\
      & $4.66 \times 10^{13} < M < 1.02\times 10^{14}$  &  $0.71 \pm 0.26$  &  $0.90 \pm 0.49$  &  $1.12 \pm 0.41$  &  $0.63 \pm 0.17$ \\
      & $M > 1.02\times 10^{14}$  &  $0.79 \pm 0.13$  &  $0.93 \pm 0.21$  &  $0.73 \pm 0.15$  &  $0.53 \pm 0.07$ \\
\hline
\end{tabular}
\end{center}
\caption{Values of the $q$-parameter, defined in Eq.~(\ref{eq:q_fudge}), obtained from $N$-body simulations for various values of
$\fnlcmb$, redshift, and mass bin.  (We take $\xi=1$ throughout)}
\label{tab:q}
\end{table}

\section{Discussion}

In this paper, we have compared semianalytic predictions for halo clustering to $N$-body simulations, in the two-field 
inflationary model from \cite{Tseliakhovich:2010kf}, in which the initial curvature fluctuation is a sum of Gaussian and 
non-Gaussian contributions.  This model is parameterized by $f_{NL}$, which corresponds to the amplitude of the 3-point function
in squeezed triangles, and a parameter $\xi$ which corresponds to the ratio of inflaton to curvaton fluctuations and boosts
the 4-point and higher functions relative to the 3-point function.  Note that curvaton models also generally predict a cubic contribution
of the form $(g_{NL} \zeta_G^3)$ to the initial curvature.  We have not considered such a term here since the peak-background
split analysis differs significantly from the $f_{NL}$ case, and defer study of the $g_{NL}$ term to future work \cite{Smith:2011ub}.

The halo bias $b(k) = P_{mh}(k)/P_{mm}(k)$ in simulation is found to agree very well
with the peak-background split prediction~(\ref{eq:bias_prediction}) on scales $k \le 0.04$ $h$~Mpc$^{-1}$,
for a range of redshifts and halo masses.
In the Gaussian case, the bias is constant in $k$ at the percent level.
In the non-Gaussian case, we find deviations from the functional form for $b(k)$ predicted by the peak-background split
which are small ($\lsim 10$\%) but statistically significant for our simulation volume.
We interpret this as agreement with the prediction, since the peak-background split is expected to break down
at the $\sim$10\% level.

We also compare the shot noise subtracted halo stochasticity parameter
\be
r(k) = \frac{P_{hh}(k) - 1/n}{P_{mm}(k)} - \left( \frac{P_{mh}(k)}{P_{mm}(k)} \right)^2
\ee
measured in simulation to semianalytic predictions.  In this case the results are more puzzling; some of the 
semianalytic predictions are confirmed and others are not.  In the Gaussian case ($\fnlcmb=0$), the halo model
makes a prediction for $r(k)$ (the leading contribution is from the 1-halo term in $P_{mh}$),
but this prediction does not consistently agree with simulation.
The excess stochasticity (relative to the halo model prediction~(\ref{eq:stochasticity_prediction})) observed in simulations
is consistent with an additive Poisson-like contribution to the halo-halo power spectrum 
(i.e.~$\Delta P_{hh}(k)$ is independent of $k$ on large scales, but can depend on redshift and halo mass).

In the non-Gaussian case,
the halo model also predicts that the stochasticity does not depend on $\fnlcmb$ (if $\xi=0$); we find that
this is true in the simulations.
The last prediction is a specific functional form~(\ref{eq:delta_rii_prediction}) for the difference stochasticity $(\Delta r_{ii})$
in stochasticity between a model with $\xi > 0$ and a Gaussian cosmology.
We find that the simulations deviate from this prediction by an overall multiplicative factor $q \approx 0.7$, 
a significant enough disagreement that it should probably be incorporated when constraining the two-parameter 
curvaton model from observations.
It is not clear whether the disagreements between theory and simulation in the Gaussian and non-Gaussian cases
are related; it is not straightforward to compare the two since the excess stochasticity appears to be ``additive'' in
the Gaussian case and ``multiplicative'' in the non-Gaussian case.

Large-scale halo clustering has emerged in the last few years as one of the most powerful probes of primordial
non-Gaussianity of one of the local types (i.e.~either $f_{NL}^{\rm local}$, $g_{NL}^{\rm local}$, or
the two-field local type considered here).  In this paper, we have confirmed qualitative predictions from
the peak-background split ansatz, showing that the peak-background split picture is very useful for relating
local-type primordial non-Gaussianity to observations.  However, detailed comparison reveals differences which
are large enough ($\approx 30$\% in this case) to be important for data analysis, highlighting the need for
simulations.  The next few years should bring a mixture of theoretical, simulation-based, and observational
work which will greatly sharpen our observational constraints on the physics of the early universe.

\subsection*{Acknowledgements}

We thank Ravi Sheth, David Spergel, and Matias Zaldarriaga for helpful discussions, and Adam Brown and Alex Dahlen
for encouraging us to pursue this project.
K.~M.~S.~is supported by a Lyman Spitzer fellowship in the Department of Astrophysical Sciences at Princeton University.
M.~L.~is supported as a Friends of the Institute for Advanced Study Member and by the NSF though AST-0807444.
Simulations in this paper were performed at the TIGRESS high performance computer center at Princeton University which is jointly 
supported by the Princeton Institute for Computational Science and Engineering and the Princeton University Office of Information Technology.

\bibliographystyle{ieeetr}
\bibliography{fnl_stochastic}

\appendix

\section{Estimators for $b$ and $r$}
\label{app:estimators}

Throughout this paper, we have given estimates of the halo bias $b_i(k) = P_{mi}(k)/P_{mm}(k)$ and 
stochasticity $r_{ii}(k) = (P_{ii}(k) - 1/n_i)/P_{mm}(k) - (P_{mi}(k)/P_{mm}(k))^2$,
with error bars that are used for parameter fitting and computing $\chi^2$ statistics.
In this appendix we describe our estimator methodology, in particular the calculation of error bars.

In principle, error bars could be assigned by running multiple $N$-body simulations and using the Monte Carlo scatter in estimates of $b$ or $r$,
but this is impractical since the ``error on the error'' would be $\sqrt{2/N_{\rm mc}}$ where $N_{\rm mc}$ is the number of $N$-body simulations, so computing error bars
with $10$\% accuracy would require running $N_{\rm mc} \approx 200$ simulations.
Therefore, an analytic prescription for the error bars is necessary.

One can see intuitively that in the limit of zero stochasiticity ($r \rightarrow 0$) and zero shot noise ($n_i \rightarrow \infty$), the
statistical errors on $b$ and $r$ should go to zero, since there will be no scatter around the mean relation $P_{ii} = b_i P_{mi} = b_i^2 P_{mm}$.
Put another way, the statistical errors should not receive contributions from sample variance, since sample variance cancels when we take
ratios of power spectra.
This will be reflected in our final expressions for the error bars (Eqs.~(\ref{eq:varb_prefinal}) and~(\ref{eq:varr_prefinal}) below), in which all terms contain
prefactors of $r$ or $(1/n)$.

In a finite volume $V$ we use the Fourier conventions
\ba
\delta(\k) &=& \int d^3\x\, \delta(\x) e^{i\k\cdot\x} \\
\delta(\x) &=& V^{-1} \sum_{\k} \delta(\k) e^{-i\k\cdot\x}
\ea
With these conventions, the infinite-volume two-point function $\langle \delta(\k) \delta(\k')^* \rangle = P(k) (2\pi)^3 \delta^3(\k-\k')$ becomes
\be
\langle \delta(\k) \delta(\k')^* \rangle = V P(k) \delta_{\k\k'}
\ee
In a $k$-bin $b$, we estimate power spectra using the estimator
\be
\hat P_{\alpha\beta} = \frac{1}{N_k V} \sum_{\k\in b} \delta_\alpha(\k)^* \delta_\beta(\k)
\ee
where $N_k = \sum_{\k\in b} 1$ is the number of Fourier modes in the $k$-bin, and indices $\alpha$, $\beta$ can denote either
the matter overdensity field or a halo mass bin $i$.

For the rest of the appendix, we fix the redshift, the $k$-bin $b$, and the halo mass bin $i$.
We will make the approximation that variation in power spectra across a single $k$-bin is not important, and use the
compressed notation $P_{\alpha\beta}=P_{\alpha\beta}(k)$.
We will also use compressed notations $n=n_i$ for the halo number density, 
$b = P_{mi}/P_{mm}$ for the bias, 
and $r = (P_{ii}-1/n_i)/P_{mm} - (P_{mi}/P_{mm})^2$ for the stochasticity.

Define estimators for halo bias and stochasticity by:
\ba
\hat b &=& \frac{\hat P_{mi}}{\hat P_{mm}}  \label{eq:bhat_def} \\
\hat r &=& \frac{1}{\hP_{mm}} \left[ \left(\frac{N_k-2}{N_k-1}\right) \hP_{ii} - \left(\frac{N_k-2}{N_k}\right) \frac{1}{n} \right]
                  - \frac{N_k-2}{N_k-1} \left( \frac{\hat P_{mi}}{\hat P_{mm}} \right)^2  \label{eq:rhat_def}
\ea
The factors of $(N_k-2)/(N_k-1)$ and $(N_k-2)/N_k$ in the second equation are {\em ad hoc} for now, but we will show 
(Eqs.~(\ref{eq:b_unbiased}),~(\ref{eq:r_unbiased}) below) that they ensure that $\hat r$ is an unbiased estimator of 
$r$ in the case where $N_k$ is not $\gg 1$.

To calculate $\Var(\hat b)$ and $\Var(\hat r)$, we will make the approximation that the matter overdensity $\delta_m$ and halo overdensity $\delta_i$
are Gaussian fields.\footnote{This assumption may appear inconsistent, since are considering non-Gaussian initial conditions, and in addition $\delta_i$ 
will be a non-Gaussian field even if the initial conditions are Gaussian.  However, we will only use the final expressions for $\Var(\hat b)$ and
$\Var(\hat r)$ on large scales ($k \le 0.04$ $h$~Mpc$^{-1}$), and on these scales the fields $\delta_m$, $\delta_i$ should be approximately Gaussian.}
We would first like to characterize the joint PDF of the power spectrum estimators $\hP_{mm}$, $\hP_{mi}$ and $\hP_{ii}$.
Let us first define normalized fields
\ba
\delta_1 &=& P_{mm}^{-1/2} \delta_m \nn \\
\delta_2 &=& \left( rP_{mm}+ \frac{1}{n} \right)^{-1/2} (\delta_i - b \delta_m)
\ea
whose power spectra are normalized to $P_{11} = P_{12} = 1$ and $P_{12} = 0$.  The two sets of power spectrum estimators are related by:
\ba
\hP_{mm} &=& P_{mm} \hP_{11} \nn \\
\hP_{mi} &=& b P_{mm} \hP_{11} + P_{mm}^{1/2} \left( rP_{mm} + \frac{1}{n} \right)^{1/2} \hP_{12} \nn \\
\hP_{ii} &=& b^2 P_{mm} \hP_{11} + 2b P_{mm}^{1/2} \left( rP_{mm} + \frac{1}{n} \right)^{1/2} \hP_{12} + \left( rP_{mm} + \frac{1}{n} \right) \hP_{22}  \label{eq:ps12}
\ea
The joint PDF of the random variables $(N_k\hP_{11})$, $(N_k\hP_{12})$, and $(N_k\hP_{22})$ is well-studied; it is known as the
Wishart distribution.
A convenient way to characterize this distribution is via the Bartlett-Cholesky decomposition, which states that:
\be
\left( \barr{cc}
  N_k\hP_{11} & N_k\hP_{12} \\
  N_k\hP_{12} & N_k\hP_{22} 
\earr \right) 
=
\left( \barr{cc}
  \chi_0^{1/2} & 0 \\
  \alpha & \chi_1^{1/2} 
\earr \right)
\left( \barr{cc}
  \chi_0^{1/2} & 0 \\
  \alpha & \chi_1^{1/2}
\earr \right)^T  \label{eq:bartlett}
\ee
where the random variable $\chi_0$ is $\chi^2$-distributed with $N_k$ degrees of freedom, $\chi_1$ is $\chi^2$-distributed with $(N_k-1)$ degrees of freedom,
$\alpha$ is distributed as a unit Gaussian, and the three variables $\chi_0$, $\chi_1$, and $\alpha$ are statistically independent.

Combining Eqs.~(\ref{eq:ps12}) and~(\ref{eq:bartlett}), we find that
\ba
\hP_{mm} &=& \frac{P_{mm} \chi_0}{N_k} \nn \\
\hP_{mi} &=& \frac{P_{mm} \chi_0}{N_k} \left[ b + \left(r + \frac{1}{nP_{mm}} \right)^{1/2} \alpha\chi_0^{-1/2} \right] \nn \\
\hP_{ii} &=& \frac{P_{mm} \chi_0}{N_k} \left[ b + \left(r + \frac{1}{nP_{mm}} \right)^{1/2} \alpha\chi_0^{-1/2} \right]^2 
               + \frac{P_{mm} \chi_1}{N_k} \left( r + \frac{1}{nP_{mm}} \right)  \label{eq:joint_pdf}
\ea
These equations, combined with the statement at the end of the previous paragraph which gives the joint PDF of $\chi_0$, $\chi_1$, and $\alpha$,
completely characterize the sampling PDF of the estimators $\hP_{mm}$, $\hP_{mi}$ and $\hP_{ii}$.

Armed with this characterization, it is easy to calculate $\Var(\hat b)$ and $\Var(\hat r)$.
We write the estimators $\hat b$ and $\hat r$ (defined in Eqs.~(\ref{eq:bhat_def}),~(\ref{eq:rhat_def}))
in terms of the variables $\chi_0$, $\chi_1$, and $\alpha$ (using Eq.~(\ref{eq:joint_pdf})),
obtaining:
\ba
\hat b &=& b + \left( r + \frac{1}{nP_{mm}} \right)^{1/2} \alpha \chi_0^{-1/2} \nn \\
\hat r &=& \frac{N_k-2}{N_k-1} \left( r + \frac{1}{nP_{mm}} \right) \frac{\chi_1}{\chi_0} - \left( \frac{N_k-2}{nP_{mm}} \right) \frac{1}{\chi_0}
\ea
It is then straightforward to calculate the mean and variance of the quantities on the right-hand side, obtaining:
\ba
\langle \hat b \rangle &=& b  \label{eq:b_unbiased} \\
\langle \hat r \rangle &=& r  \label{eq:r_unbiased} \\
\Var( \hat b ) &=& \frac{1}{N_k-2} \left(r + \frac{1}{nP_{mm}} \right)  \label{eq:varb_prefinal} \\
\Var( \hat r ) &=& \frac{2}{N_k-4} \left( r^2 \right) + \frac{2(N_k-2)}{(N_k-1)(N_k-4)} \left( r + \frac{1}{nP_{mm}} \right)^2  \label{eq:varr_prefinal}
\ea
For these calculations the expectation value $\langle \chi^m \rangle = 2^m \Gamma(m+N/2) / \Gamma(N/2)$, where $\chi$ is $\chi^2$-distributed
with $N$ degrees of freedom, is useful.

This calculation is exact even in the case where $N_k$ is not $\gg 1$, and this level of precision appears to be necessary,
e.g.~we find that if terms of order $(1/N_k)$ in the variance are neglected, then a few jackknife tests in Tab.~\ref{tab:jackknife} below fail.
We also note that both estimators are unbiased (i.e.~$\langle \hat b \rangle = b$ and $\langle \hat r \rangle = r$, justifying the factors
of $(N_k-1)$ and $(N_k-2)$ in the definition~(\ref{eq:rhat_def}) of $\hat r$, which were {\em ad hoc} until
now.  (The estimators would be biased if these factors were omitted.)

There is one final wrinkle: in order to apply these expressions for $\Var(\hat b)$ and $\Var(\hat r)$, we need to know the stochasticity $r$.
In this paper we do not propose a general model for stochasticity, finding for example that the nonzero stochasticity seen in simulations for 
Gaussian initial conditions is not fit well by the halo model prediction (\S\ref{ssec:halo_stochasticity}), and makes a non-negligible contribution to the
estimator variance.
Therefore, we infer the stochasticity directly from the simulation itself, by making the approximation
\be
r \approx \frac{\hP_{ii} - 1/n}{\hP_{mm}} - \left( \frac{\hP_{mi}}{\hP_{mm}} \right)^2
\ee
on the right-hand sides of Eqs.~(\ref{eq:varb_prefinal}),~(\ref{eq:varr_prefinal}).

Putting the results of this appendix together, our final estimates for $\Var(\hat b)$ and $\Var(\hat r)$ are given by:
\ba
\Var(\hat b) & \approx & \frac{1}{N_k-2} \left[ \left( \frac{\hP_{ii}}{\hP_{mm}} \right) - \left( \frac{\hP_{mi}}{\hP_{mm}} \right)^2 \right] \\
\Var(\hat r) & \approx & 
    \frac{2}{N_k-4} \left[ \left( \frac{\hP_{ii}-1/n}{\hP_{mm}} \right) - \left( \frac{\hP_{mi}}{\hP_{mm}} \right)^2 \right]^2
  + \frac{2(N_k-2)}{(N_k-1)(N_k-4)} \left[ \left( \frac{\hP_{ii}}{\hP_{mm}} \right) - \left( \frac{\hP_{mi}}{\hP_{mm}} \right)^2 \right]^2 \nn
\ea
We have used these expressions to assign error bars throughout this paper, e.g.~in Figs.~\ref{fig:bias} and~\ref{fig:stochasticity}.

We conclude this appendix with an end-to-end test of our estimates for $\Var(\hat b)$ and $\Var(\hat r)$.
If we run two independent $N$-body simulations with the same values of $\fnlcmb$ and $\xi$, then the differences
$(\hat b - \hat b')$ between bias estimates, and differences $(\hat r - \hat r')$ between stochasticity
estimates, should be consistent with zero.
This jackknife test is intended to check correctness of the error bars without requiring a model for the expected
values of the bias and stochasticity.

More precisely, for each halo mass bin $i$, we define a $\chi^2$ statistic by summing over $k$-bins $b$:
\be
\chi^2 = \sum_b \frac{\hat r_{ii}(b) - \hat r'_{ii}(b)}{\Var(\hat r_{ii}(b)) + \Var(\hat r'_{ii}(b))}
\ee
and analogously with the stochasticity estimator $\hat r$ replaced by the bias estimator $\hat b$.
Results from the jackknife tests are shown in Tab.~\ref{tab:jackknife}.
The results appear consistent with $\chi^2$ statistics, indicating that our error estimates are accurate.
(The most anomalous $\chi^2$ value in the table is 27.6 with 14 degrees of freedom, corresponding to a $p$-value
of 1.6\%.  Since there are 100 entries in the table, an entry which is anomalous at this level is expected.)

\begin{table}[!h]
\begin{center}
\setlength{\tabcolsep}{3pt}
\begin{tabular}{|c|c|c|c|c|c|c|}
\hline & & \multicolumn{5}{c|}{$\chi^2$ for $\hat b_{mi}$ jackknife test ($N_{\rm dof}=14$)} \\ \cline{3-7}
  & Mass range ($h^{-1} M_\odot$) & \small{$f_{NL}=0$} & \small{$f_{NL}=500$} & \small{$f_{NL}=-500$} & \small{$f_{NL}=500$} & \small{$f_{NL}=-500$}  \\
  &   & \small{$\xi=0$} & \small{$\xi=0$} & \small{$\xi=0$} & \small{$\xi=1$} & \small{$\xi=1$} \\ \hline\hline
$z=2$ & $M > 1.15 \times 10^{13}$ & 23.4 & 27.6 & 16.7 & 14.3 & 11.1 \\
\hline $z=1$ & $1.15 \times 10^{13} < M < 2.32\times 10^{13}$ & 16.5 & 13.9 & 7.4 & 16.1 & 15.1 \\
      & $M > 2.32 \times 10^{13}$ & 16.0 & 14.4 & 24.1 & 13.1 & 6.9 \\
\hline $z=0.5$ & $1.15 \times 10^{13} < M < 2.32\times 10^{13}$ & 17.1 & 19.1 & 22.2 & 10.1 & 9.8 \\
        & $2.32 \times 10^{13} < M < 4.66\times 10^{13}$ & 9.4 & 7.0 & 9.4 & 19.5 & 20.9 \\
        & $M > 4.66 \times 10^{13}$ & 19.3 & 24.0 & 12.3 & 8.7 & 7.9 \\
\hline $z=0$ & $1.15 \times 10^{13} < M < 2.32\times 10^{13}$ & 16.7 & 13.1 & 11.4 & 14.2 & 22.5 \\
      & $2.32 \times 10^{13} < M < 4.66\times 10^{13}$ & 25.2 & 14.5 & 16.5 & 20.0 & 8.7 \\
      & $4.66 \times 10^{13} < M < 1.02\times 10^{14}$ & 12.3 & 14.5 & 11.4 & 13.6 & 12.1 \\
      & $M > 1.02\times 10^{14}$ & 21.6 & 10.9 & 18.1 & 9.5 & 12.1 \\
\hline
\end{tabular}
\vskip 1cm
\begin{tabular}{|c|c|c|c|c|c|c|}
\hline & & \multicolumn{5}{c|}{$\chi^2$ for $\hat r_{ii}$ jackknife test ($N_{\rm dof}=14$)} \\ \cline{3-7}
  & Mass range ($h^{-1} M_\odot$) & \small{$f_{NL}=0$} & \small{$f_{NL}=500$} & \small{$f_{NL}=-500$} & \small{$f_{NL}=500$} & \small{$f_{NL}=-500$}  \\
  &   & \small{$\xi=0$} & \small{$\xi=0$} & \small{$\xi=0$} & \small{$\xi=1$} & \small{$\xi=1$} \\ \hline\hline
$z=2$ & $M > 1.15 \times 10^{13}$ & 16.9 & 14.0 & 19.5 & 9.6 & 7.9 \\
\hline $z=1$ & $1.15 \times 10^{13} < M < 2.32\times 10^{13}$ & 12.7 & 10.7 & 6.8 & 10.9 & 10.7 \\
      & $M > 2.32 \times 10^{13}$ & 7.3 & 10.2 & 13.2 & 9.3 & 8.8 \\
\hline $z=0.5$ & $1.15 \times 10^{13} < M < 2.32\times 10^{13}$ & 16.8 & 11.8 & 9.0 & 8.2 & 18.9 \\
        & $2.32 \times 10^{13} < M < 4.66\times 10^{13}$ & 9.4 & 9.3 & 5.2 & 14.2 & 21.8 \\
        & $M > 4.66 \times 10^{13}$ & 6.7 & 12.1 & 8.3 & 12.0 & 9.7 \\
\hline $z=0$ & $1.15 \times 10^{13} < M < 2.32\times 10^{13}$ & 9.8 & 10.8 & 15.7 & 8.3 & 9.9 \\
      & $2.32 \times 10^{13} < M < 4.66\times 10^{13}$ & 7.2 & 13.3 & 9.0 & 12.0 & 7.1 \\
      & $4.66 \times 10^{13} < M < 1.02\times 10^{14}$ & 14.9 & 7.8 & 9.8 & 9.7 & 7.6 \\
      & $M > 1.02\times 10^{14}$ & 24.4 & 20.0 & 18.3 & 15.9 & 21.9 \\
\hline
\end{tabular}
\end{center}
\caption{Jackknife tests for the bias estimator $\hat b_{mi}$ and stochasticity estimator $\hat r_{ii}$.  The $\chi^2$ values are
consistent with statistical expectations, showing that the variances $\Var(\hat b_{mi})$ and $\Var(\hat r_{ii})$ have been
correctly estimated.}
\label{tab:jackknife}
\end{table}

\end{document}